%% file: ms_x1.tex
\documentclass[twocolumn,twocolappendix]{aastex63}

\hypersetup{linkcolor=blue,citecolor=blue,filecolor=magenta,urlcolor=cyan}

\shorttitle{LOS Velocity Selected \oVI\ CGM Measurements}
\shortauthors{Ho, Martin, \& Schaye}

\usepackage{natbib}
\usepackage{bm}
\usepackage{chngcntr}

\begin{document}

\input setdef.tex

\title{How identifying circumgalactic gas
by line-of-sight velocity instead of the location in 3D space
affects \oVI\ measurements}


\correspondingauthor{Stephanie Ho}
\email{shho@physics.tamu.edu}

\author[0000-0002-9607-7365]{Stephanie H. Ho}
\affiliation{George P.~and Cynthia Woods Mitchell Institute for Fundamental Physics and Astronomy, Texas A\&M University, College Station, TX 77843-4242, USA}
\affiliation{Department of Physics and Astronomy, Texas A\&M University, College Station, TX 77843-4242, USA}

\author[0000-0001-9189-7818]{Crystal L. Martin}
\affiliation{Department of Physics, University of California, Santa Barbara, CA 93106, USA}

\author[0000-0002-0668-5560]{Joop Schaye}
\affiliation{Leiden Observatory, Leiden University, P.O. Box 9513, 2300 RA, Leiden, The Netherlands}



\begin{abstract}

The high incidence rate of the \oVIdb\ absorption 
around low-redshift, $\sim$\lstar\ star-forming galaxies 
has generated interest in studies of the circumgalactic medium.
We use the high-resolution \EAGLE\ cosmological simulation
to analyze the circumgalactic \oVI\ gas 
around $z\approx0.3$ star-forming galaxies.
Motivated by the limitation that 
observations do not reveal 
where the gas lies along the line-of-sight,
we compare the \oVI\ measurements 
produced by gas within fixed distances around galaxies
and by gas selected using line-of-sight velocity cuts
commonly adopted by observers.
We show that gas selected by a velocity cut
of $\pm300$\kms\ or $\pm500$\kms\ 
produces a higher \oVI\ column density,
a flatter column density profile,
and a higher covering fraction
compared to gas within
one, two, or three times the virial radius (\rvir) of galaxies.
The discrepancy increases with impact parameter
and worsens for lower mass galaxies.
For example, 
compared to the gas within 2\rvir,
identifying the gas using velocity cuts of 200-500\kms\
increases the \oVI\ column density by
0.2 dex (0.1 dex) at 1\rvir\
to over 0.75 dex (0.7 dex) at $\approx2$\rvir\
for galaxies with stellar masses of
$10^{9}$-$10^{9.5}$\msununit\ ($10^{10}$-$10^{10.5}$\msununit).
We furthermore estimate that excluding \oVI\ outside \rvir\ 
decreases the circumgalactic oxygen mass measured by 
\citet{Tumlinson2011} by over 50\%.
Our results demonstrate that gas 
at large line-of-sight separations
but selected by conventional velocity windows
has significant effects on the \oVI\ measurements
and may not be observationally distinguishable
from gas near the galaxies.

\end{abstract}


\section{Introduction}
\label{sec:intro}

Absorption-line spectroscopy has
revealed the substantial reservoir 
of baryons and metals surrounding galaxies, 
known as the circumgalactic medium 
(CGM, \citealt{Tumlinson2017}, and references therein).
The CGM extends to at least the virial radius \rvir\
and regulates the interplay between the 
gas accretion onto galaxies and the feedback from massive stars.
The ubiquitous detection of the
\oVIdb\ absorption in sightlines
intersecting the CGM of $\sim$\lstar\ star-forming galaxies
in contrast to the rare \oVI\ detection  
around quiescent galaxies 
has drawn particular attention \citep{Tumlinson2011}.
This observed ``\oVI\ bimodality'' 
possibly indicates a link between \oVI\ and 
ongoing star formation.
This link may however be indirect.  
The absence of detectable \oVI\ may indicate lower CGM mass
fractions, which in turn may result from 
active galactic nucleus (AGN) feedback 
\citep{Davies2020,Nelson2018}, or higher
virial temperatures and hence halo masses 
\citep{Oppenheimer2016}.

Dedicated observational efforts have 
characterized the circumgalactic \oVI\ properties
and explored its dependence on galaxy properties
and its relationship with 
the low-ionization-state (LIS) absorbers (e.g., \mgII, \siII).
The $\sim$\lstar\ star-forming galaxies 
do not only have a higher \oVI\ incidence rate and column density
than quiescent galaxies \citep{Tumlinson2011}
but also compared with dwarf galaxies 
\citep{Prochaska2011,Johnson2015,Johnson2017}
and massive luminous red galaxies \citep{Zahedy2019}.
This suggests that the strength of the \oVI\ absorption
depends on star formation and/or the galaxy mass.
The \oVI\ column density decreases with 
the sightline impact parameter,
but the decline is less steep \citep{Tumlinson2011,Stern2018}
compared to the LIS absorbers 
\citep{Chen2010,Huang2021}.
Even though \oVI\ systems have broader line profiles 
than the LIS counterparts \citep{Werk2016,Nielsen2017},
most \oVI\ absorption components
have matching LIS components with similar Doppler shifts
\citep{Werk2016}.
However, unlike the 
tentative evidence for stronger and broader 
LIS absorption detected
near the galaxy minor axes compared to that near the
galaxy major axes \citep{Kacprzak2012,Nielsen2015}, 
a result typically 
attributed to galactic outflows
\citep{Martin2019,Schroetter2019},
\oVI\ absorption is kinematically uniform at 
all azimuthal angles \citep{Nielsen2017}.\footnote{
    The azimuthal angle is the angle between 
    the galaxy major-axis and the line joining the 
    sightline and the center of the galaxy.
    }
The \oVI\ Doppler shifts measured in major-axis sightlines
do not correlate with disk rotation \citep{Kacprzak2019},
whereas the LIS gas corotates with the galaxy
\citep{Ho2017,Zabl2019,HoMartin2020}.
These similarities and differences 
in the low ions and \oVI\ properties
highlight the complexity of the multiphase CGM.

One challenge for interpreting 
the observed circumgalactic absorption, is that
absorption-line measurements do not reveal where 
the gas lies along the line-of-sight (LOS).
Observers typically associate the absorbing gas 
with the galaxy at the smallest
impact parameter from the sightline 
that has a comparable redshift,
i.e., within a preset 
line-of-sight velocity window from the 
galaxy systemic velocity. 
However, 
the gas potentially resides relatively
far away from the galaxy in 3D space and 
may therefore have no direct
physical relation with the galaxy
(see \citealt{Ho2020} for a study of this issue
for \mgII\ absorption).
Faint galaxies responsible for the absorption
may also remain undetected
until deeper imaging and spectroscopy 
of the galaxy field becomes available,
and incorrectly associating the gas to
another bright galaxy in the field 
would alter the interpretation of the 
origin of the detected gas.
These uncertainties in identifying the
gas associated with target galaxies
lead to possible errors in 
interpreting the CGM properties
from absorption-line measurements.

The ambiguity of the relative location between the gas
and the galaxies
does not pose a problem for CGM analyses 
using large volume hydrodynamic simulations,
though zoom-in simulations may underestimate 
the projection effects.
Simulations that reproduce
the radial profiles of the column density of LIS gas 
\citep{Ford2016,Oppenheimer2018lowion}
often underestimate the \oVI\ column density
around $\sim$\lstar\ galaxies by about a factor of two
\citep[e.g.,][]{Hummels2013,Oppenheimer2016,
Gutcke2017,Suresh2017,Marra2020}.
This problem could potentially be resolved
by fossil AGN proximity zones
\citep{OppenheimerSchaye2013,Oppenheimer2018agn},
black hole feedback \citep{Nelson2018},
by including cosmic ray physics \citep{Ji2020},
or by changing the model for the
UV background \citep{Appleby2021}.
While simulations reproduced the 
\oVI\ bimodality observed in the star-forming and quiescent
galaxy sample from 
Tumlinson \etal\ \citep{Oppenheimer2016,Nelson2018},
Oppenheimer \etal\ suggested that 
the observed bimodality was due to
the higher halo mass of the quiescent galaxy sample;
the virial temperature of the $\sim$\lstar\ star-forming
galaxies ($\sim10^{5.5}$ K)
coincides with the narrow temperature range
at which the \oVI\ fraction peaks
in a collisionally ionized plasma.

This paper focuses on the circumgalactic \oVI\ gas 
using the high-resolution \EAGLE\ simulation
\citep{Schaye2015,Crain2015}.  
\EAGLE\ broadly reproduces many galaxy observables,
e.g., the galaxy stellar mass function \citep{Schaye2015},
the evolution of galaxy masses \citep{Furlong2015}, 
sizes \citep{Furlong2017}, 
colors \citep{Trayford2015,Trayford2017}, 
and gas contents \citep{Lagos2015,Bahe2016,Crain2017}.  
Although \EAGLE\ was not calibrated to match 
the observed CGM properties,
the simulation shows
broad agreement with absorption-line statistics 
for both \hI\ and metal ions \citep{Rahmati2015,Rahmati2016,Turner2016,Turner2017,
Oppenheimer2018lowion}.
For example, \EAGLE\ reproduces
the anticorrelation between
the covering fraction and impact parameter
of low ions \citep{Oppenheimer2018lowion} 
and the observed \oVI\ bimodality
in low-redshift galaxies \citep{Oppenheimer2016}.

In this paper, 
we analyze the \oVI\ gas around low-redshift galaxies
and focus on how selecting the gas around galaxies
by LOS velocity instead of 3D distance
affects the \oVI\ measurements 
and the interpretations of the \oVI\ properties of galaxies.
We present the paper as follows.
Section~\ref{sec:sample_simulation}
describes the \EAGLE\ simulation. 
In Section~\ref{sec:profiles},
we present the \oVI\ measurements
and contrast the results from 
gas selected using different fixed radii and 
LOS velocity windows.
We discuss the observational consequences and conclude 
in Section~\ref{sec:discussion}.
Throughout this paper,
we use the flat $\Lambda$CDM cosmology
with $(\Omega_m,\Omega_\Lambda,h) = (0.307,0.693,0.6777)$
adopted by \EAGLE\ from \citet{Planck2014}.
We prefix comoving and proper (i.e., physical) 
length units with ``c'' and ``p'', respectively, 
e.g., cMpc and pkpc.

\section{Galaxy Selection in the \EAGLE\ Simulation
and Column Density Maps}
\label{sec:sample_simulation}

\subsection{Simulation Overview and Galaxy Selection}
\label{ssec:simulation}

The \EAGLE\ simulation suite consists of a large
number of cosmological hydrodynamical simulations
with different resolutions, cosmological volumes,
and physical models \citep{Schaye2015,Crain2015,McAlpine2016}.
\EAGLE\ was performed using a modified version 
of the $N$-Body Tree-PM 
smoothed particle hydrodynamics (SPH) code \GADGET\
(last described in \citealt{Springel2005})
with a new hydrodynamics solver
\citep{Schaller2015}.
Unresolved physical processes are
captured by state-of-the-art subgrid models,
including radiative cooling and photoheating, 
star formation, stellar evolution and enrichment, 
stellar feedback, 
and black hole growth and AGN feedback.
\citet{Schaye2015} introduced a reference model;
the subgrid model parameters
for energy feedback from stars and accreting black holes 
were calibrated to reproduce 
the $z\approx0$ galaxy stellar mass function, 
the sizes of disk galaxies,
and the amplitude of the
galaxy-central black hole mass relation.

\EAGLE\ defines galaxies using the
\texttt{SUBFIND} algorithm \citep{Springel2001,Dolag2009}.
Briefly, the friends-of-friends (FoF) algorithm
connects dark matter particles to the same halo
if the particle separation is below 0.2 times
the average particle separation.
Baryons are linked to the same halo (if it exists)
as their closest dark matter particle.
In each FoF halo,
\texttt{SUBFIND} identifies self-bound overdense regions
as subhalos.
Each subhalo represents a galaxy.
The central galaxy is defined as the subhalo with 
the particle at the lowest gravitational potential,
and the location of this particle represents 
the center of the central galaxy.

In this work, we use the Recal-L0025N0752 simulation
with a box size of 25 cMpc.\footnote{
    The ``Recal'' model was calibrated to 
    the same $z\approx0$ galaxy properties 
    as the reference model, 
    but subgrid parameters for stellar and AGN feedback 
    were modified and recalibrated
    as a consequence of the higher resolution 
    compared to the default (intermediate) resolution runs.
    }
This simulation has 
a dark matter particle mass of $1.21\times10^6$ \msununit, 
an initial baryonic particle mass of $2.26\times10^5$ \msununit,
and a Plummer-equivalent gravitational softening length of
0.35 pkpc at the low-redshift we study here.
This simulation run has
8 (2) times better mass (spatial)
resolution compared to the default \EAGLE\ intermediate-resolution
runs (e.g., Ref-L0100N1504).
We analyze the $z=0.271$ snapshot output;\footnote{
    Particle data snapshots can be downloaded from
    \texttt{\url{http://icc.dur.ac.uk/Eagle/database.php}}
    }
this redshift is comparable to the redshifts of targeted galaxies
in circumgalactic \oVI\ observations
\citep[e.g.,][]{Tumlinson2011,Johnson2015,Johnson2017,Kacprzak2019,Beckett2021}.
Because \EAGLE\ applied periodic boundary conditions,
the maximum LOS separation is half of the box size
and corresponds to a velocity difference
of 767 \kms\ (physical) at $z=0.271$.

\begin{figure}[thb]
    \centering
    \includegraphics[width=1.0\linewidth]{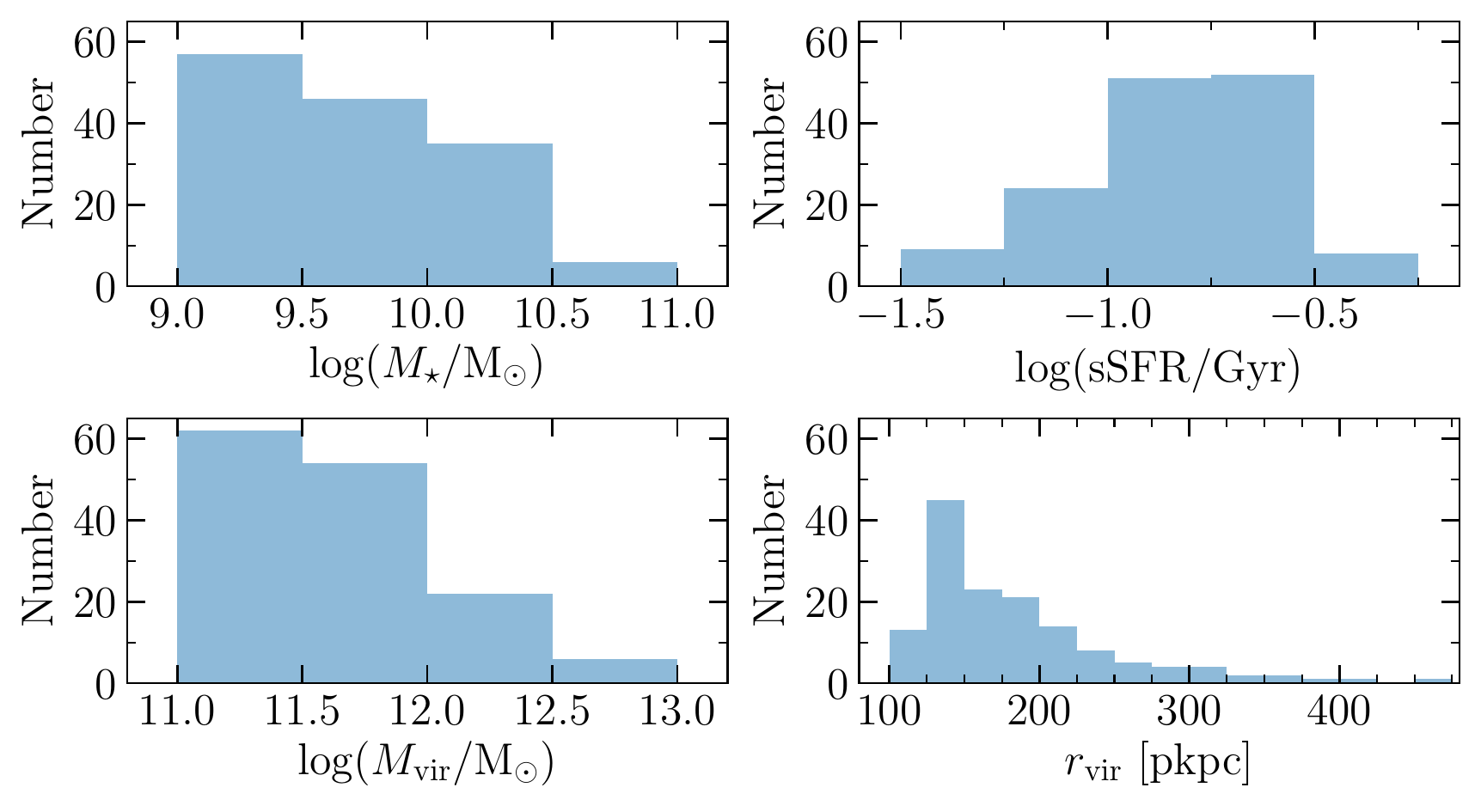}
    \caption{
        Distributions of 
        stellar mass \mstar, 
        specific star formation rate (sSFR),
        halo virial mass \mvir, and virial radius \rvir\
        of the selected central, star-forming galaxies.
        The sample consists of 144 star-forming galaxies.
        }
    \label{fig:hist_galprop} 
\end{figure}

We select central, star-forming galaxies 
with stellar masses (\mstar)  between 
$10^9$ to $10^{11}$ \msununit.
The galaxy stellar mass is defined as the 
total mass of star particles associated with the
subhalo within 30 pkpc (in 3D)
from the galaxy center \citep{Schaye2015}.
Following \citet{Ho2020}, 
galaxies are classified as star-forming
if they lie above the 
dividing line between star-forming and quiescent galaxies
on the SFR-\mstar\ plane
as defined by \citet{Moustakas2013},
who fitted a redshift-dependent relation 
to separate star-forming and quiescent galaxies
using $\sim$120,000 spectroscopically observed galaxies in
the PRism MUlti-object Survey
(also see Figure~1 of \citealt{Ho2020}).
Our sample consists of 144 star-forming galaxies; 
Figure~\ref{fig:hist_galprop} shows the distributions
of the galaxy stellar mass, specific star formation rate,
halo virial mass, and virial radius of the sample.\footnote{
    The virial radius \rvir\ is the 
    radius enclosing an average density of
    $\Delta_\mathrm{vir} \rho_c(z)$,
    where $\rho_c(z)$ represents the critical density 
    at redshift $z$.
    The overdensity $\Delta_\mathrm{vir}$ follows
    the top-hat spherical collapse calculation
    in \citet{BryanNorman1998}.  
    The halo virial mass is the total mass enclosed
    within the sphere of radius \rvir.
    }

\subsection{Column Density Maps}
\label{ssec:2dmap}

We project galaxies along the Z-axis in the simulation box
and produce the \oVI\ (and \mgII) column density maps.
Calculating the ionic column density requires
the element abundance and the ion fraction, 
i.e., 
the number of atoms in each ionization state divided by 
the total number of atoms of the element in the gas phase.
For \oVI\
we use the ion fraction tables from \citet{Bertone2010a,Bertone2010b},
who adopted the UV/X-ray background 
by \citet{HaardtMadau2001} and
computed the tables under the same assumptions
as the gas cooling in the \EAGLE\ runs.
For \mgII\
we obtain the ion fraction using the fiducial model  
in \citet{PloeckingerSchaye2020},
whose calculations 
use the UV/X-ray background by
\citet{FaucherGiguere2020}\footnote{
    \citet{PloeckingerSchaye2020} 
    modified the $z>3$ UV/X-ray background in
    \citet{FaucherGiguere2020}  
    to improve the self-consistency of the 
    treatment of attenuation before \hI\ and \heII\
    reionization (their Appendix B).
    This modification is irrelevant to this work
    at low redshift.}
and include the effects of depletion of metals onto dust grains
and self-shielding (also see \citealt{Ho2020}).  
The \mgII\ fraction will be slightly different
compared to that used for computing the cooling rates 
during \EAGLE\ run, because \EAGLE\ used a 
different UV background model and did not include self-shielding.
However, we do not expect the difference to be significant,
because magnesium is not an important coolant \citep{Wiersma2009}.  
We interpolate the ion fraction tables
in redshift, log temperature, and log density
to obtain the \oVI\ and \mgII\ ion fractions per SPH particle.
Then we multiply the ion fraction by the particle mass and 
the element abundance to calculate the number of ions per particle.  
We obtain the column density by summing 
the total number of ions through a gas column and divide that
by the cross-sectional area of the column, 
during which the spatial distribution of gas (i.e., the ion) 
of each SPH particle is modeled 
by the same $C^2$ \citet{Wendland1995} kernel used for 
the hydrodynamics calculations in the \EAGLE\ simulations 
(also see \citealt{Wijers2019}).
For our galaxies, 
\mgII\ traces $T$$\sim$$10^4$ K gas and is photoionized,
whereas collisionally ionized \oVI\ ($T$$\approx$$10^{5.5}$K)
dominates the inner radius.
Photoionized \oVI\ becomes increasingly important at larger radii
especially for lower mass galaxies
(with $T\lesssim10^5$K and density 
$n_H\lesssim10^{-4.5}$\percmcube).
This agrees with the radial and galaxy mass dependence 
of \oVI\ ionization 
shown in recent work from zoom-in simulations
\citep{RocaFabrega2019,Strawn2021}
and the ``low-pressure'' \oVI\ scenario presented
in \citet{Stern2018}, for which cool, photoionized 
\oVI\ exists beyond the accretion shock.   
Detailed discussion on the ionization mechanism is 
beyond the scope of this paper 
(see \citealt{Oppenheimer2016}, \citealt{Rahmati2016}, and 
\citealt{Wijers2020}).

We use two approaches to select the gas around galaxies
while making the column density maps.
Our first approach selects only gas within fixed 3D radii 
of 1, 2, and 3\rvir\ from each galaxy center.
This method excludes gas further away from the galaxy but 
appearing nearby because of the 2D projection.
The second approach selects gas 
within a fixed LOS velocity difference \deltavlos\
from the galaxy systemic velocity.\footnote{\
    The galaxy systemic velocity
    is the mass-weighted velocity
    of all particles (stars, gas, dark matter, and black hole)
    associated with the galaxy at the $z=0.271$ snapshot.}
We use \deltavlos\ = 300\kms\ and 500\kms,
both of which are commonly adopted in observational 
analyses to identify absorption systems 
associated with the target galaxies
(e.g., \citealt{Chen2010,Werk2016}).
On the column density maps, 
each pixel shows the total gas column 
summed along the path enclosed
by the sphere of a fixed radius 
or the LOS velocity window.
Each pixel has an area of (1.25 pkpc)$^2$ or (0.005 \rvir)$^2$.

\begin{figure}[tbh]
    \centering 
    \includegraphics[width=1.0\linewidth]{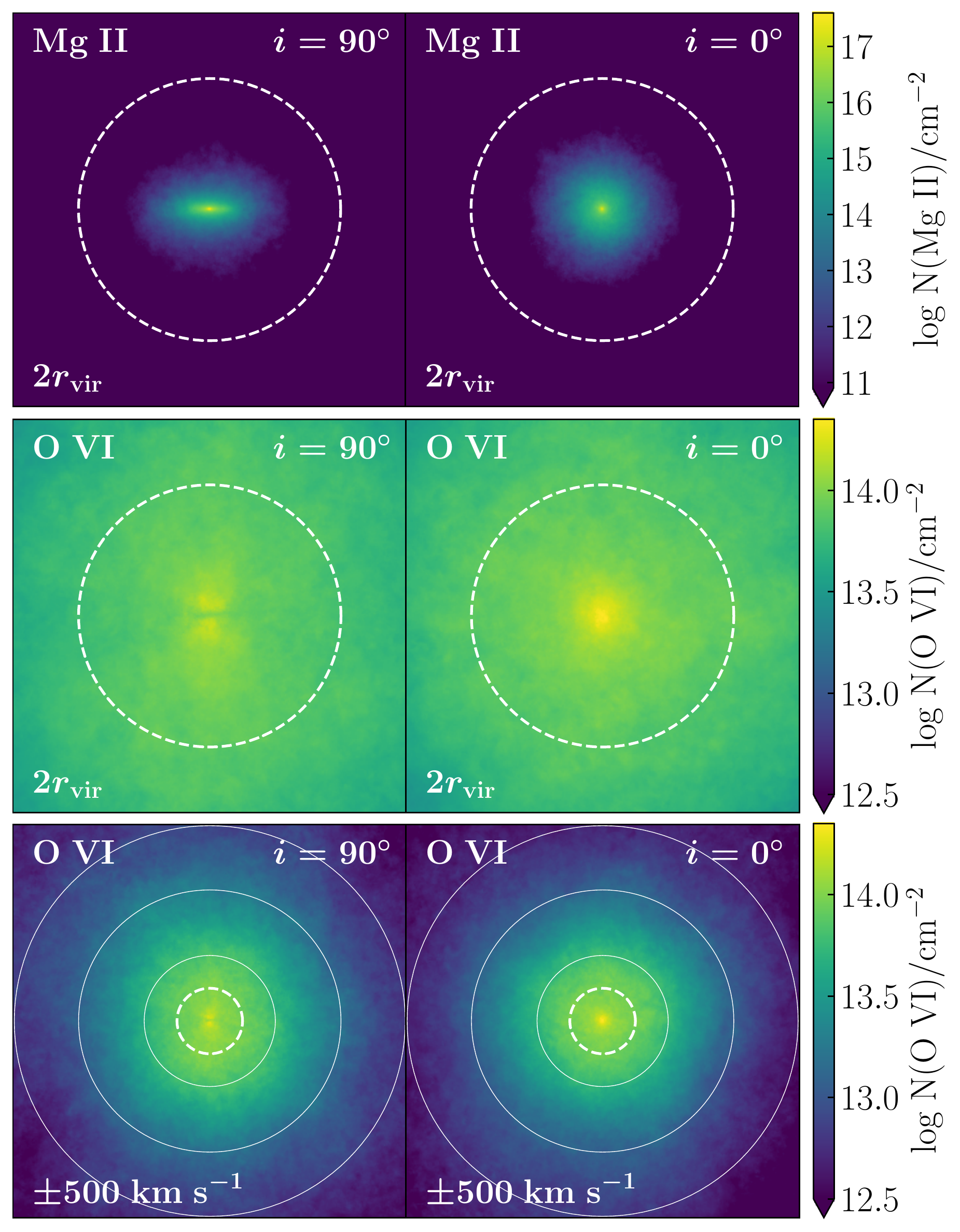}
    \caption{
        Median \mgII\ \textit{(top)} and
        \oVI\ \textit{(middle and bottom)} column density
        around all star-forming central galaxies in the sample.
        The median stellar mass and halo mass of the stacked galaxies
        are $10^{9.7}$ \msununit\ and $10^{11.6}$ \msununit, 
        respectively.
        In the top and middle rows, only the gas within 
        2\rvir\ of individual galaxies is included.  
        Each white dashed circle marks a radius of 0.5\rvir.
        The bottom row zooms out and shows the 
        \oVI\ gas distribution at larger scale.  
        The concentric solid circles
        mark the radii of 1, 2, and 3\rvir.
        The gas around individual galaxies is selected 
        using the LOS velocity window of \deltavlos\ = 500\kms.  
        The left and right columns show the edge-on ($i=90^\circ$)
        and face-on projections ($i=0^\circ$), respectively.
        The galaxy orientation
        is defined using the stellar angular momentum vector
        (as in \citealt{Ho2020}).
        The \mgII\ gas
        distribution is concentrated near the galaxy center,
        whereas the \oVI\ gas extends to a larger radius.
        }
    \label{fig:coldens_map} 
\end{figure}

As an illustration,
the maps in Figure~\ref{fig:coldens_map} show 
the median \mgII\ (top) and \oVI\ (middle and bottom)
column density for the stacks of all star-forming galaxies.
For this Figure only,
individual galaxies are projected edge-on ($i=90$\deg, left) or
face-on ($i=0$\deg, right) before they are stacked together.
The \oVI\ gas clearly extends to a larger radius than \mgII,
which is concentrated within 0.4\rvir\ around the galaxy center.
The \oVI\ and \mgII\ gas also show different morphologies,
for which only \mgII\ but not \oVI\ shows 
an axisymmetric structure.
We defer the discussion of \oVI\ morphology  
and its dependence on inclination and azimuthal angles
to a future paper (Ho \etal, in prep.)
and refer readers
to \citet{Ho2020} for the \mgII\ analysis.

\section{\oVI\ Column Density and Covering Fraction Measurements}
\label{sec:profiles}

The high \oVI\ incidence rate out to 
at least 150 pkpc ($\approx$\rvir)
from $\sim$\lstar\ star-forming galaxies suggest that 
\oVI\ extends to this radius or further
\citep{Tumlinson2011}.
In this Section, we use \EAGLE\ to demonstrate  
how selecting the gas within
different fixed LOS separations
around the target galaxies (1, 2, and 3\rvir)
and different LOS velocity windows 
(\deltavlos\ = 300\kms\ and 500\kms)
changes the measurements for \oVI\ 
column density (Section~\ref{ssec:coldens})
and covering fraction (Section~\ref{ssec:covfrac}).

\subsection{Column Density Profile}
\label{ssec:coldens}

\begin{figure*}[tbh]
    \centering
    \includegraphics[width=1.0\linewidth]{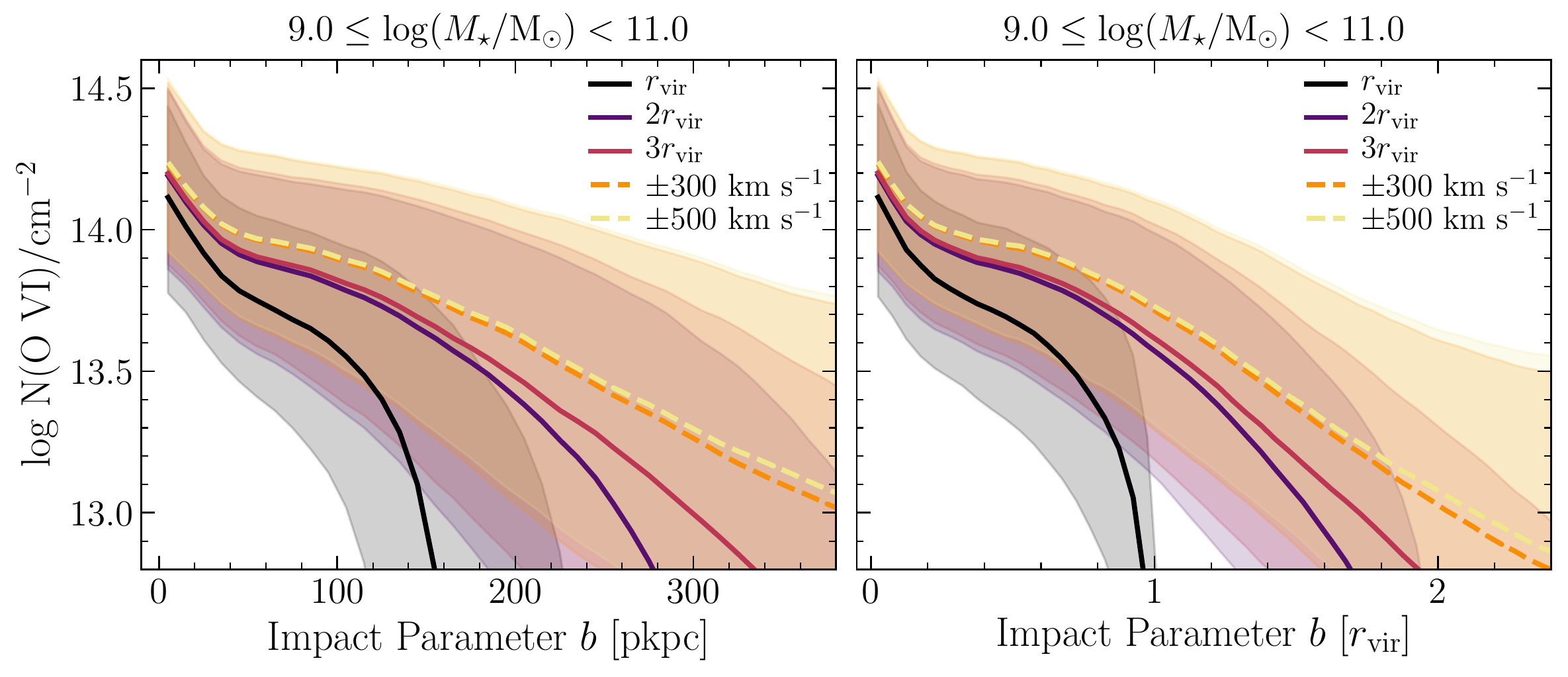}
    \caption{
        \oVI\ column density as a function of 
        impact parameter \textit{(left)} and 
        that normalized by halo virial radius \textit{(right)}.
        Different colors show the column density
        measured from the gas  
        within 1, 2, or 3 \rvir\ of individual galaxies 
        or within a LOS velocity window
        of $\pm300$\kms\ or $\pm500$\kms\
        from the galaxy systemic velocity.
        Each line shows the median from all pixels around 
        all star-forming galaxies, 
        and each shaded region of the same color encloses the 
        16$^{\rm{th}}$ and 84$^{\rm{th}}$ percentiles.
        At all impact parameters, 
        selecting gas using a fixed LOS velocity 
        window produces 
        a higher column density and
        a flatter column density profile compared to 
        that from gas even within 3\rvir.
        }
    \label{fig:o6_coldens} 
\end{figure*}

We create the column density profiles 
for stacks of randomly oriented galaxies 
using all the pixels 
from individual galaxies in each stack.
For the stack of all star-forming galaxies, 
Figure~\ref{fig:o6_coldens} shows the 
\oVI\ column density 
as a function of impact parameter $b$ (left) 
and that normalized by \rvir\ (right).  
Each line shows the median column density 
from gas selected using different criteria.

First, the \oVI\ gas clearly extends beyond \rvir.
The sharp column density drop-off from 
gas selected within \rvir\ (black) is not observed 
from gas within 2\rvir\ (purple) or 3\rvir\ (magenta).  
Second, 
at a fixed impact parameter,
gas selected by a larger radius around galaxies
produces a higher column density.
For impact parameters $b\approx$ \rvir, 
the \oVI\ absorption is dominated by gas at $\approx2$\rvir\
(see also \citealt{Oppenheimer2016} and \citealt{Wijers2020}).
Although selecting the gas within 2\rvir\ and 3\rvir\
produces a significantly smaller column density difference 
compared to that from gas 
within 1\rvir\ and larger radii,
this still implies 
a non-negligible amount of \oVI\ beyond 2\rvir.
This extended \oVI\ distribution is 
in stark contrast to the centrally concentrated 
\mgII\ distribution (e.g., Figure~\ref{fig:coldens_map}).

More importantly, 
selecting gas using a LOS velocity window \deltavlos\ 
of 300\kms\ (orange) or 500\kms\ (yellow)
produces a higher \oVI\ column density 
at all impact parameters,
even compared to that from gas physically within 3\rvir.
The difference increases with impact parameter.
Consequently, 
gas selected by \deltavlos\ produces a flatter 
\oVI\ column density profile 
compared to that from gas within a fixed radius.
Altogether, 
our results imply that 
even at a small impact parameter
($b < 150$ pkpc or $< 1$\rvir),
selecting gas using a fixed LOS velocity window
includes a large contribution 
from \oVI\ that is relatively far away from the galaxies.

\begin{figure*}[htb]
    \centering
    \includegraphics[width=1.0\linewidth]{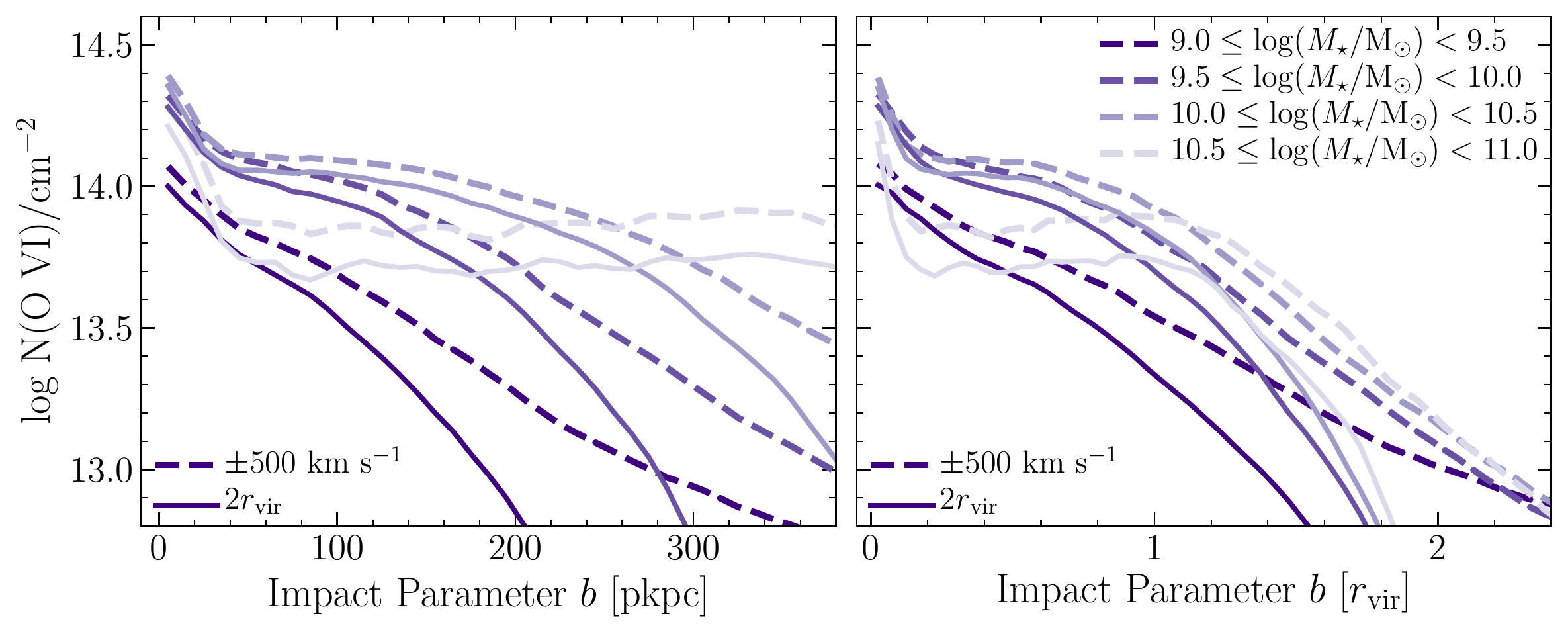}
    \newline
    \includegraphics[width=1.0\linewidth]{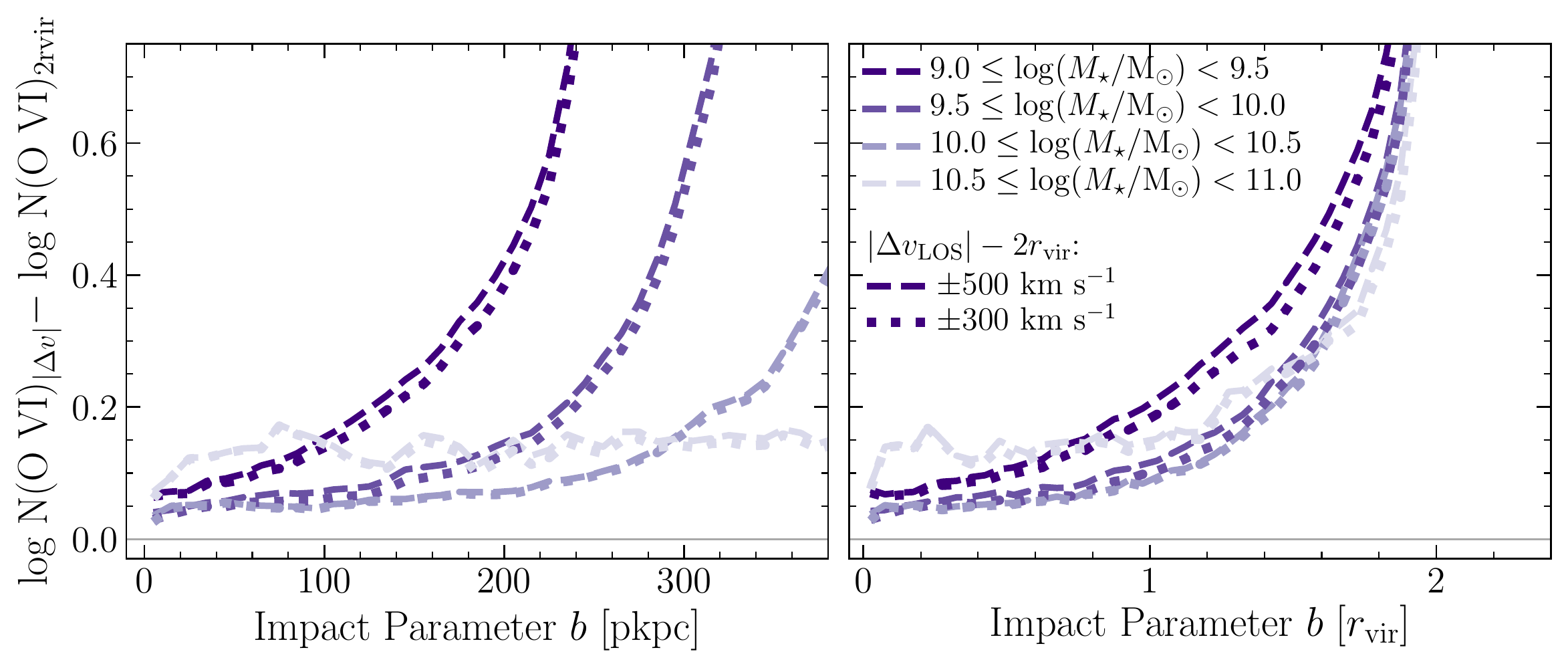}
    \caption{
        Median \oVI\ column density around 
        star-forming galaxies 
        in different stellar mass bins as a function of 
        impact parameter \textit{(left)} and 
        that normalized by halo virial radius \textit{(right)}.
        The top row shows the \oVI\ column density profiles 
        from gas inside 2\rvir\ (solid lines) 
        and gas selected by
        a LOS velocity window of \deltavlos\ = 500\kms\ 
        (dashed lines).
        The bottom row shows the difference in
        the median \oVI\ column density profiles
        between gas selected using \deltavlos\
        and gas within 2\rvir.
        The dashed (dotted) lines 
        represent the results
        from \deltavlos\ = 500\kms\ (300 \kms).
        In all panels, darker (lighter) lines show the results
        for galaxies in the lower (higher) mass bins.
        In all mass bins, 
        gas within \deltavlos\ produces 
        a higher \oVI\ column density than that 
        from gas within 2\rvir,
        and the difference increases with impact parameter.
        }
    \label{fig:o6_coldens_4mass} 
\end{figure*}

Figure~\ref{fig:o6_coldens_4mass} shows
the median \oVI\ column density profiles for galaxies 
in different stellar mass bins.
The top row shows the profiles
from gas physically within 2\rvir\ (solid lines)
and gas selected by the LOS velocity window 
of \deltavlos\ = 500\kms\ (dashed lines).
Darker (lighter) lines represent the measurements 
around lower (higher) mass galaxies.
In all mass bins, similar to Figure~\ref{fig:o6_coldens},
gas within \deltavlos\ = 500\kms\ produces a 
higher column density than that from gas within 2\rvir\
at a fixed impact parameter. 
To illustrate how the column density difference 
varies with impact parameter and galaxy stellar mass,
the bottom row shows 
the difference of the median column density profiles.
The dashed (dotted) lines show 
the results from gas within 
\deltavlos\ of 500\kms\ (300\kms)
compared to that from gas inside 2\rvir.
These plots clearly show that 
the column density difference increases with impact parameter.
Then at a fixed impact parameter,
the darker lines generally lie above the lighter lines.
This indicates that 
the difference in the \oVI\ column density is typically 
larger for a lower mass galaxy.
For example, at $b = 150$ pkpc or $\approx$1\rvir,
the $9.0 \leq$ \logmstarmsun\ $< 9.5$  galaxies
detect a difference of about 0.25 dex, 
whereas the difference is around 0.1 dex 
for the $10.0 \leq$ \logmstarmsun\ $< 10.5$ galaxies.
In other words,
by selecting the gas using  the $\pm300$\kms\ or 
$\pm500$\kms\ window instead of a fixed radius of 2\rvir,
the measured \oVI\ column density increases  
by 25\% - 80\% at an impact parameter 
of 1\rvir.

\subsection{Covering Fraction}
\label{ssec:covfrac}

Because selecting gas using a commonly adopted
LOS velocity window
produces higher \oVI\ column density
than that from gas within a fixed radius of $\sim$\rvir\
around galaxies,
gas outside the fixed radius 
but selected by \deltavlos\
will elevate the \oVI\ detection rate 
for a fixed column density threshold.
To illustrate this effect, 
we calculate the \oVI\ covering fraction
as a function of impact parameter.
First, we bin the pixels of all galaxies in each stack
by the impact parameter $b$ or \brvir.
Then, in each bin,
we count the number of pixels with \oVI\ column density 
above the threshold
and divide that number by the 
total number of pixels in the bin.

\begin{figure*}[tbh]
    \centering
    \includegraphics[width=1.0\linewidth]{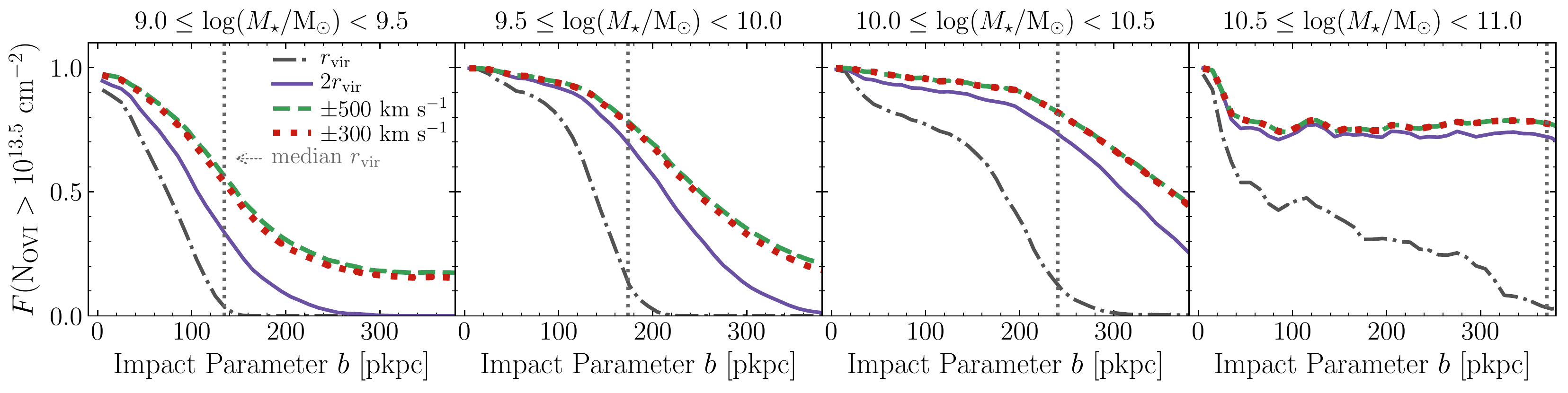}
    \newline
    \includegraphics[width=1.0\linewidth]{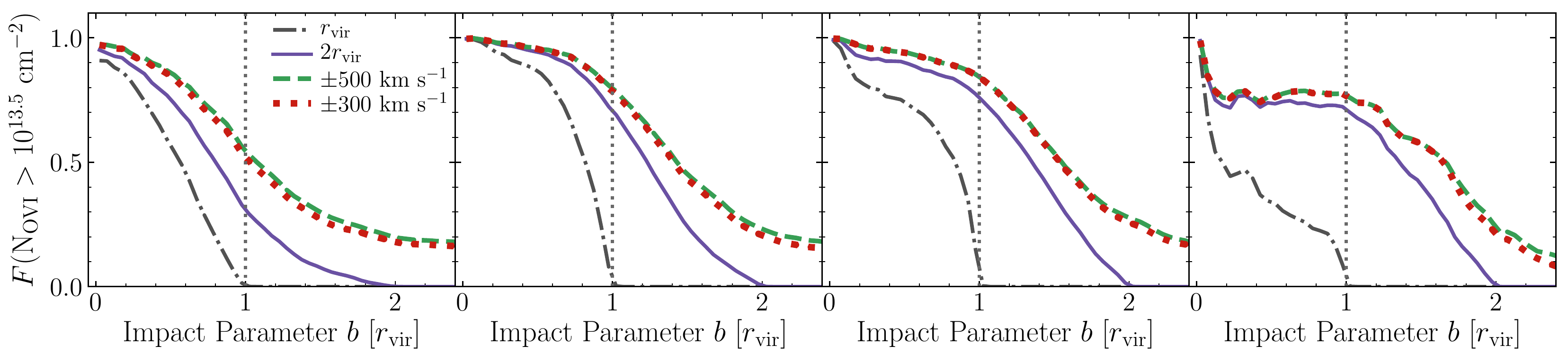}
    \caption{
        \oVI\ covering fraction for 
        star-forming galaxies
        as a function of impact parameter \textit{(top)} and 
        that normalized by the halo virial radius \textit{(bottom)}.
        The adopted detection threshold
        is \logNovi/\percmsq = 13.5.
        Different columns show the results for galaxies 
        in different stellar mass bins,
        and different line styles show the results
        from gas selected using different criteria.
        The vertical dotted lines 
        show the median \rvir\ of the galaxies 
        in the sample.
        At a fixed impact parameter,
        gas selected by \deltavlos\ always produces
        a higher covering fraction 
        than gas within 2\rvir,
        and lower mass galaxies show a larger difference.
        }
    \label{fig:o6_detfrac} 
\end{figure*}

Figure~\ref{fig:o6_detfrac} shows the 
\oVI\ covering fraction
for star-forming galaxies 
in different stellar mass bins (columns).
The top and bottom rows show the fraction 
as a function of $b$ and \brvir, respectively.
We adopt a detection threshold
of \logNovi/\percmsq = 13.5,
which is comparable to the 
lowest \oVI\ column density detected 
in quasar sightlines around low-redshift galaxies
\citep[e.g.,][]{Tumlinson2011,Johnson2015}.
In all mass bins, 
selecting gas within 
500\kms\ (green dashed lines) and
\deltavlos\ = 300\kms\ (red dotted lines)
produces comparable covering fractions
with a difference no greater than 0.03.
In contrast, 
selecting gas using either value of \deltavlos\
produces a higher covering fraction than gas 
within 1\rvir\ or 2\rvir,
and the difference is more prominent 
at larger impact parameters and for lower mass galaxies.
For instance, at $b = $ 1\rvir\ (vertical dotted lines),
selecting the gas using 1\rvir\ produces a zero covering
fraction (by definition)\footnote{
    In the top row of Figure~\ref{fig:o6_detfrac},
    gas selected within 1\rvir\ (dot-dashed curves)
    does not produce a zero covering fraction 
    at the vertical dotted lines,
    because the vertical dotted lines only 
    show the median \rvir\ of galaxies 
    in individual mass bins.
    },
but the fraction is clearly above zero 
when we adopt other selection criteria.
If we select the gas using \deltavlos\ instead of 2\rvir,
then the \oVI\ covering fraction around
$9 \leq$ \logmstarmsun\ $< 9.5$
($10 \leq$ \logmstarmsun\ $< 10.5$) galaxies
increases from 0.3 to 0.55 (0.75 to 0.85),
implying an $\approx$85\% (15\%) increase.

\section{Discussion and Conclusion}
\label{sec:discussion}

In the last decade, the ubiquitous \oVI\ absorption
detected around $\sim$\lstar\ star-forming galaxies
out to a large impact parameter of $\gtrsim$150 pkpc has
generated interest in CGM studies.
A challenge for understanding and interpreting 
circumgalactic absorption is that the measurements
do not reveal where the detected gas lies along the LOS.
Hence, observers typically identify absorption systems
associated with the target galaxies using a LOS velocity cut 
around the galaxy systemic velocity
(e.g., $\pm300$\kms\ and $\pm500$\kms).
In this paper, we used 
the high-resolution \EAGLE\ (25 cMpc)$^3$ simulation
and analyzed the \oVI\ gas around 
$z\approx0.25$, star-forming galaxies.
We demonstrated that the \oVI\ column density
and detection rate 
depend on whether we include
gas around galaxies within a fixed 3D radius (1, 2, and 3\rvir)
or gas selected using a LOS velocity window.

We showed that 
selecting gas using the 
commonly adopted LOS velocity windows 
of \deltavlos\ = 300\kms\ or 500\kms\
always produces a higher \oVI\ column density
and a flatter column density profile
compared to gas within a fixed radius
(Figure~\ref{fig:o6_coldens}).
The column density discrepancy increases with impact parameter
and generally worsens for lower mass galaxies 
(Figure~\ref{fig:o6_coldens_4mass}).
For example, when we compared the column density
measured from gas within \deltavlos\ and 2\rvir, 
the difference increases from
0.2 dex (0.1 dex) at 1\rvir\
to over 0.75 dex (0.7 dex) at $\approx2$\rvir\
for $9.0 \leq$ \logmstarmsun\ $< 9.5$ 
($10.0 \leq$ \logmstarmsun\ $< 10.5$) galaxies. 
Because selecting gas using the \deltavlos\ criterion
increases the measured \oVI\ column density,
it also raises the \oVI\ detection rate 
for a fixed detection threshold,
i.e., the covering fraction (Figure~\ref{fig:o6_detfrac}).

\subsection{Examples of how the enhanced \oVI\ column density
affects the interpretation of observational measurements}
\label{ssec:examples}

We use the COS-Halos dataset \citep{Tumlinson2011} 
to demonstrate how the elevated \oVI\ column density
from the LOS velocity selection affects the 
\oVI\ mass estimation deduced from observational data.
Their sample consists of 30 (12) sightlines 
within $b = 150$ pkpc, i.e., $\lesssim$\rvir,
of $z\approx0.2$, $\sim$\lstar\ star-forming (quiescent) 
galaxies, and 27 (4) of the sightlines have detected \oVI.
By projecting a circular region 
of radius $R = 150$ pkpc on the sky,
they estimated the total mass of \oVI\ around 
star-forming galaxies using
$M_\mathrm{O\ VI} = \pi R^2  m_\mathrm{O} \kappa \langle N_\mathrm{O\ VI}\rangle$,
where $m_\mathrm{O}$ is the oxygen atomic mass, 
$\langle N_\mathrm{O\ VI} \rangle$ is the mean \oVI\ column density, 
and $\kappa$ is the covering fraction.
They measured the 
$\log \langle N_\mathrm{O\ VI} \rangle/\mathrm{cm}^{-2}$
as 14.7, 14.6, and 14.5 in radial bins $R$ of 
0-50 pkpc, 50-100 pkpc, and 100-150 pkpc, respectively,
but they adopted 
$\log \langle N_\mathrm{O\ VI} \rangle/\mathrm{cm}^{-2} = 14.5$
to obtain an \oVI\ mass lower limit.\footnote{
    Adopting 
    $\log \langle N_\mathrm{O\ VI} \rangle/\mathrm{cm}^{-2} = 14.5$
    also avoids the calculated \oVI\ and oxygen masses
    being skewed by sightlines 
    with large \logNovi\ at small impact parameters.}
The resultant \oVI\ and oxygen mass are $2.4\times10^6$ \msununit\
and $1.2\times10^7(0.2/f_\mathrm{O\ VI})$ \msununit, respectively,
where $f_\mathrm{O\ VI}$ is the \oVI\ ionization fraction.
Because their sightlines lie within 150 pkpc, this 
calculation is often interpreted as the total \oVI\ 
and oxygen mass within a volume of R $<$ 150 pkpc, 
i.e., $\lesssim$\rvir, of the star-forming galaxies.

\begin{figure*}[htb]
    \centering
    \includegraphics[width=1.0\linewidth]{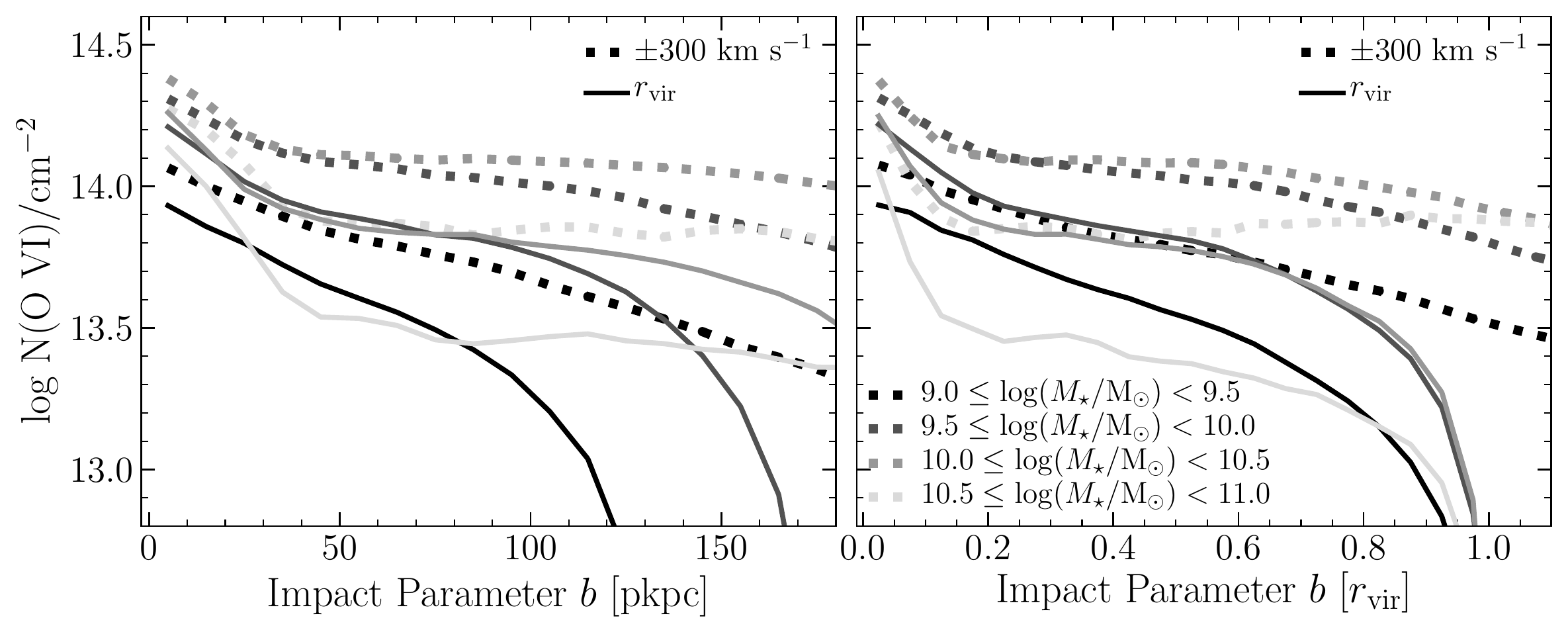}
    \caption{
        Median \oVI\ column density around 
        star-forming galaxies 
        in different stellar mass bins as a function of 
        impact parameter \textit{(left)} and 
        that normalized by halo virial radius \textit{(right)}.
        This figure is analogous to the top row of 
        Figure~\ref{fig:o6_coldens_4mass}, 
        but this figure shows the profiles 
        measured from gas within \rvir\ (solid lines) 
        and the LOS velocity window of \deltavlos\ = 300\kms\ 
        (dotted lines).
        Such selections are comparable to the COS-Halos
        dataset in \citet{Tumlinson2011};
        the sightlines lie within 150 pkpc ($\lesssim$\rvir)
        of the selected galaxies, 
        and the Doppler shifts of most \oVI\ velocity components 
        lie within $\pm300$\kms\
        of the galaxy systemic velocities.
        }
    \label{fig:o6_coldens_4mass_1r} 
\end{figure*}

However, our analysis with \EAGLE\ demonstrates that 
the LOS velocity window 
selects a non-negligible amount of \oVI\ gas outside \rvir.
Hence, we use the difference in \oVI\ column density  
measured from gas within \rvir\ and \deltavlos\
to correct the observed \logNovi\ values
and recalculate the \oVI\ mass within \rvir\ of 
the star-forming galaxies.
Because \citet{Tumlinson2011} showed that most \oVI\ velocity 
components have Doppler shifts below $\pm300$\kms,
we correct the observed column densities from
\deltavlos\ of 300\kms\ to within \rvir\ 
based on the stellar mass of each galaxy
and the sightline impact parameter.  
We note that our calculations and conclusions
remain unchanged if we use \deltavlos\ = 200\kms\
instead (as further discussed in Section~\ref{ssec:small_vlos}).
Figure~\ref{fig:o6_coldens_4mass_1r} is analogous to 
the top row of Figure~\ref{fig:o6_coldens_4mass},
but the \oVI\ column density profiles
are measured from gas within \rvir\ (solid) and 
\deltavlos\ = 300\kms\ (dotted)
around \EAGLE\ star-forming galaxies.
For the 27 observed sightlines with \oVI\ detected
in \citet{Tumlinson2011},
the average \logNovi\ correction is $-0.28$ dex, 
and the corrected 
$\log \langle N_\mathrm{O\ VI} \rangle/\mathrm{cm}^{-2}$
in 0-50 pkpc, 50-100 pkpc, and 100-150 pkpc 
are 14.47, 14.33, and 14.13, respectively.
Following Tumlinson et al.,
we calculate the mass lower limit using the 
smallest $\log \langle N_\mathrm{O\ VI} \rangle$.
The 0.37 dex decrease in 
$\log \langle N_\mathrm{O\ VI} \rangle/\mathrm{cm}^{-2}$
(from 14.5 to 14.13)
corresponds to a decrease of 57\%
for the \oVI\ and oxygen masses.
Hence, this reduces the baryon budget estimated
for the warm-hot CGM by more than 50\%.

\begin{figure}[htb]
    \centering
    \includegraphics[width=1.0\linewidth]{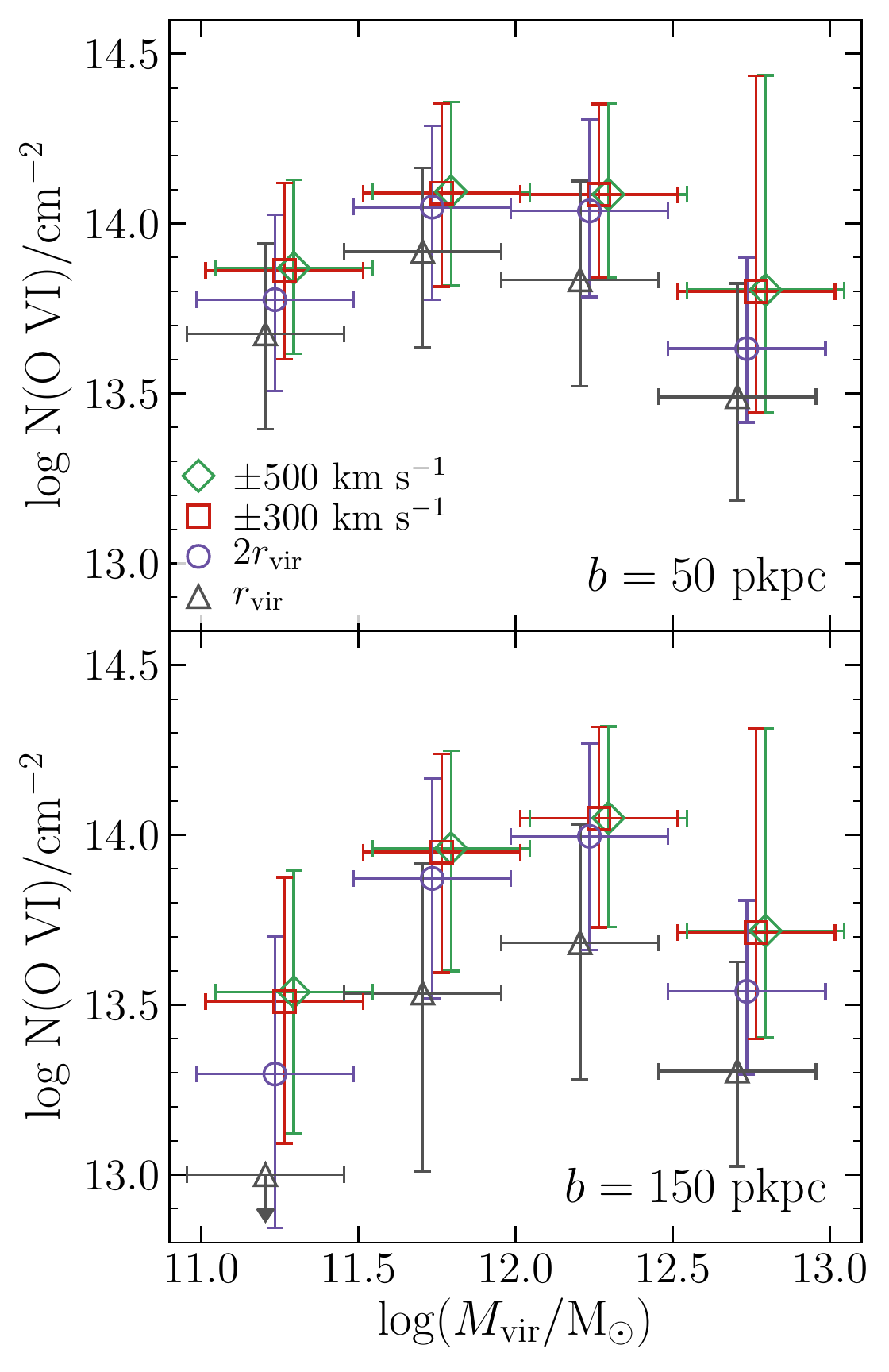}
    \caption{
        \oVI\ column density around 
        star-forming galaxies as a function of halo mass
        at impact parameters of 
        50 pkpc (top) and 150 pkpc (bottom).
        Different markers show the median \oVI\ column densities 
        measured from gas selected using different criteria.
        The error bars show the 16$^{\rm{th}}$ and 84$^{\rm{th}}$ 
        percentiles.
        Points are offset along the horizontal axis 
        to avoid the overlapping of error bars.
        At $b$ = 150 pkpc, 
        the median \oVI\ column density
        of galaxies with halo masses of
        $10^{11.0-11.5}$ \msununit\ is below the axis limit
        (downward arrow).
        Galaxies with halo masses above $10^{12.5}$ \msununit\
        have a lower \oVI\ column density than those with
        halo masses of $10^{11.5-12.5}$ \msununit.
        In all mass bins, gas within \rvir\ always produces a
        lower \oVI\ column density compared to gas selected
        by the LOS velocity windows.
        }
    \label{fig:o6_coldens_mhalo} 
\end{figure}

Detailed discussion of the rare \oVI\ detection 
around quiescent galaxies 
and the ``\oVI\ bimodality'' in the COS-Halos sample 
is beyond the scope of this paper.
However, it is worth noting that \citet{Oppenheimer2016}
suggested that this ``bimodality'' does not imply 
a causal link between sSFR and \oVI\ column density
but reflects the higher halo mass of the quiescent galaxy sample.
In fact, even with our simulated star-forming galaxies only,
those with halo masses above $10^{12.5}$ \msununit\
have a lower \logNovi\ than those slightly less massive.
As an illustration,
Figure~\ref{fig:o6_coldens_mhalo} shows
the \oVI\ column density as a function
of halo mass at fixed impact parameters
of 50 pkpc (top) and 150 pkpc (bottom).
Different markers represent the results from
gas selected using different criteria.
Regardless of the selection method we use,
galaxies with halo mass above $10^{12.5}$ \msununit\
have lower \oVI\ column densites than
those around lower mass halos of $10^{11.5-12.5}$ \msununit.
This also agrees with the halo mass dependence 
of \oVI\ column density shown in \citet{Wijers2020}; 
low-redshift \EAGLE\ galaxies 
with halo masses of $\sim10^{12}$ \msununit\
have the highest \oVI\ column density compared to 
both higher and lower mass halos (see their Figure~8).
Altogether, this demonstrates the decline of \oVI\ column density
at the high mass end as emphasized in \citet{Oppenheimer2016}
while interpreting the COS-Halos measurements.

As another example for the \oVI\ mass calculation, 
we use the \citet{Johnson2017} sample
of 18 star-forming dwarf galaxies
with stellar masses of $\sim10^{8-9}$ \msununit.
The authors estimated the \oVI\ mass within \rvir\ = 90 pkpc 
using the mean \oVI\ column density measured from
the two sightlines at the smallest impact parameters;
these two sightlines detected \oVI\ of 
\logNovi$/\mathrm{cm}^{-2}$ = 14.10 and 14.17 
at 0.23\rvir\ and 0.26\rvir, respectively.
Because our analysis did not include \EAGLE\ galaxies 
with stellar masses below $10^9$ \msununit,
we estimate the column density correction 
using the results from the \EAGLE\ galaxies 
with the closest stellar masses ($10^{9-9.5}$ \msununit).
The estimated correction is conservative;
the column density discrepancy measured from gas 
within a fixed radius and a LOS velocity window 
worsens with decreasing stellar mass
(Figures~\ref{fig:o6_coldens_4mass} and 
\ref{fig:o6_coldens_4mass_1r}).
The \logNovi\ correction from \deltavlos\ = 300\kms\ to \rvir\
for the two sightlines are $-0.16$ dex and $-0.17$ dex.  
As a result, the mean \logNovi, and hence the
estimated \oVI\ mass, decreases by over 30\%.\footnote{
    We obtain a smaller \oVI\ mass correction for the dwarf galaxy
    sample in \citet{Johnson2017} than the $\sim$\lstar\ galaxy
    sample in \citet{Tumlinson2011}.
    This is because \citet{Johnson2017} obtained 
    $\log \langle N_\mathrm{O\ VI} \rangle$
    from sightlines at small impact parameters of 
    0.23\rvir\ and 0.26\rvir, whereas 
    \citet{Tumlinson2011} used sightlines at
    larger impact parameters of 
    $b \lesssim 150$ pkpc, i.e, $\lesssim$\rvir.
    However, 
    we note that
    the correction estimated for the Johnson et al.~sample
    is conservative, because we estimate the correction
    from \EAGLE\ galaxies an order of magnitude more massive
    than the observed dwarf galaxies.
    }

Because \oVI\ column density scales with the \oVI\ mass,
the difference in \oVI\ column density between 
gas selected using a fixed radius versus a LOS velocity window
demonstrates the difference in the O VI mass included.
Even at small impact parameters, 
when observers use a LOS velocity window to identify the 
\oVI\ gas associated with the target galaxies,
a non-negligible amount, possibly even dominant
amount of the detected \oVI,
actually resides at large physical distances 
(e.g., $>$\rvir; Figures~\ref{fig:o6_coldens_4mass_1r} 
and \ref{fig:o6_coldens_mhalo}).
Not only does the gas at large distances increase
the \oVI\ detection rate and column density,
this gas potentially weakens any \oVI\ kinematic signature
produced by gas close to the galaxy.
While we defer the \oVI\ kinematic analysis
to a future paper (Ho \etal, in prep.),
a similar analysis of \mgII\ showed that
even though the \mgII\ gas corotates with 
galaxies out to at least 0.5\rvir,
the gas far away ($>$ \rvir) but selected by the LOS velocity window
makes observers less likely to conclude that
gas at $\gtrsim0.25$\rvir\ is corotating \citep{Ho2020}.
This result from \mgII\ illustrates 
that gas at large distances can ``contaminate''
the measurements and mask the ``true''
circumgalactic gas properties.

\subsection{Comments on the 
radial extent of \oVI\ gas and 
the small LOS velocity differences}
\label{ssec:small_vlos}

\begin{figure*}[tbh]
    \centering
    \includegraphics[width=1.0\linewidth]{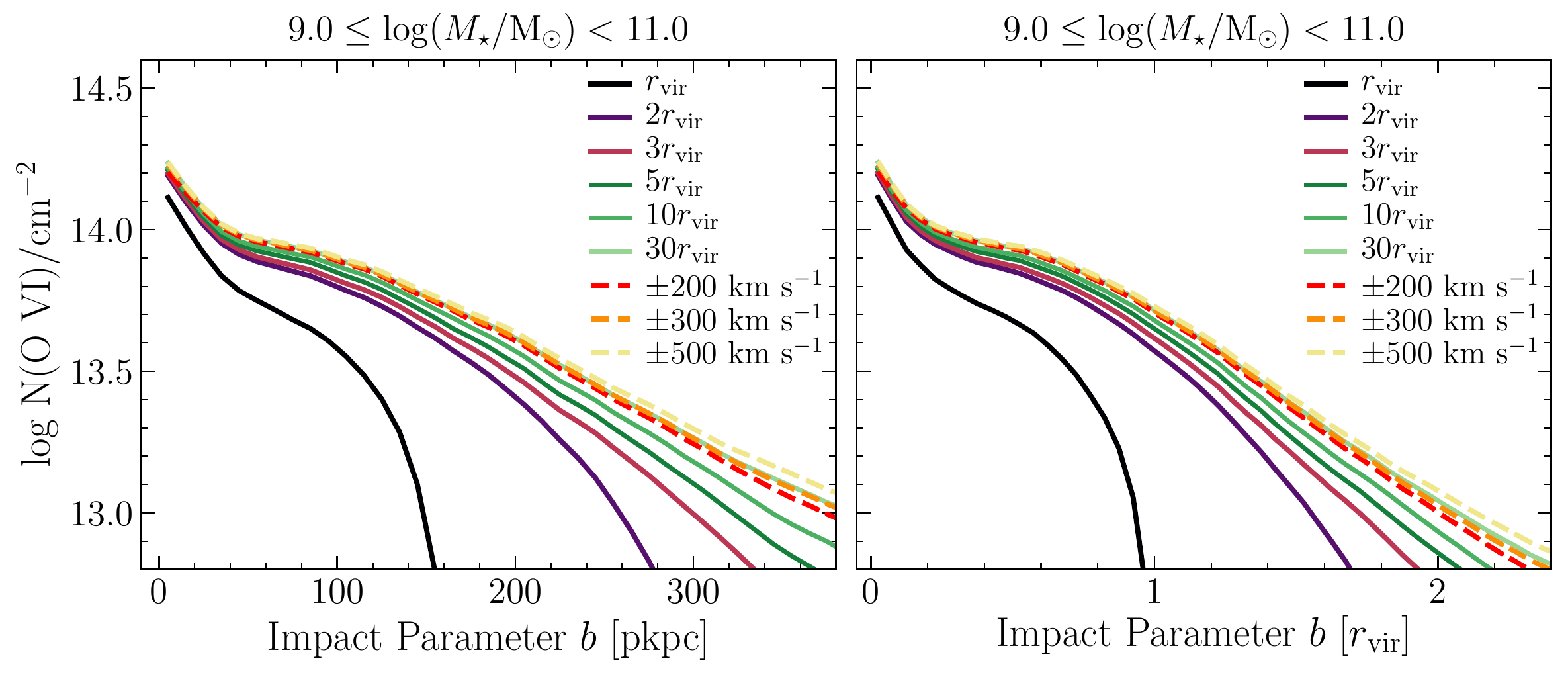}
    \caption{
        Median \oVI\ column density as a function of 
        impact parameter \textit{(left)} and 
        that normalized by halo virial radius \textit{(right)}.
        This figure is similar to Figure~\ref{fig:o6_coldens}
        but with additional column density profiles
        measured from gas within 5, 10, and 30\rvir\ 
        of individual galaxies
        (greenish solid)
        and within a LOS velocity window of $\pm200$\kms\
        (red dashed) from the galaxy systemic velocity.
        Gas within 30\rvir\ (light green solid)
        of individual galaxies has to be included to
        produce a median \oVI\ column density profile
        comparable to those 
        measured from gas selected by the LOS velocity window of 
        $\pm200$\kms, $\pm300$\kms\ or $\pm500$\kms.
        }
    \label{fig:o6_coldens_extr} 
\end{figure*}

While observers typically adopt
the $\pm300$\kms\ or $\pm500$\kms\ window
to search for absorption systems
associated with the target galaxies,
most detected \oVI\ systems have 
Doppler shifts within $\pm200$\kms\ from 
the galaxy systemic velocity
\citep[][private communication]{Tumlinson2011,Zahedy2019}.
This however is also true for simulations, and
when we repeat our analysis with
the narrower window of $\pm200$\kms,
our conclusions remain unchanged.  
Figure~\ref{fig:o6_coldens_extr} illustrates this point;
especially at small impact parameters ($\lesssim150$ pkpc or \rvir),
the \oVI\ column density profiles measured from 
gas within $\pm200$\kms\ (red dashed),
$\pm300$\kms\ (orange dashed), and $\pm500$\kms\ (yellow dashed)
largely overlap.
This implies that even out to larger scale, 
i.e., beyond 1-3\rvir\ of the galaxies 
and into the intergalactic scale, 
the \oVI\ gas still has a small LOS velocity
relative to the galaxies.

Figure~\ref{fig:o6_coldens_extr} shows that
in order to  measure an \oVI\ column density
comparable to that from gas selected by 
the LOS velocity windows of $\pm300$\kms\ (orange dashed)
and $\pm500$\kms\ (yellow dashed), 
we have to include the gas within 
30\rvir\ (light green solid) around the galaxies.
In fact, nearly 100\% of the \oVI\ gas within 30\rvir\
have LOS velocities within $\pm300$\kms\ and $\pm500$\kms\
(not shown).
Given that the median \rvir\ of our galaxy sample is 160 pkpc;
30\rvir\ corresponds to about 4.8 pMpc
and a Hubble flow velocity of 380\kms. 
This Figure clearly demonstrates that \oVI\ extends 
far beyond 1-3\rvir\
and to tens of \rvir\ at $\sim$pMpc scale.
The Hubble flow velocity of 380\kms\ also explains
why the gas at such a large LOS separation is still included
in our velocity windows; 
unless the \oVI\ resides in massive clusters with 
peculiar velocities of $\sim1000$\kms,
the \oVI\ at $\sim$pMpc scale moving with the Hubble flow
will have LOS velocity comparable to or smaller than
$\pm300$\kms\ and $\pm500$\kms.

The large spatial extent of \oVI\ gas around galaxies
and the small \oVI\ velocity along LOS
agree with results from 
previous observational (and simulation)
\oVI--galaxy cross-correlation analyses.
\citet{Prochaska2019} used the CASBaH survey  
with thousands of $0.12<z<0.75$ galaxies in nine quasar fields 
and showed that while the \oVI\ covering fraction
($\geq 10^{13.5}$\percmsq)
declines with impact parameter, 
there still exists an excess in the covering fraction 
out to $\approx$8 pMpc compared to the expectation
from random \oVI\ incidence.\footnote{
    The rate of random \oVI\ incidence is estimated 
    by surveying many sightlines in blind surveys
    and calculating the number of \oVI\ absorbers per
    redshift interval.}
This implies an \oVI--galaxy clustering out to 
this length scale,
and hence, it is not surprising that 
we find \oVI\ around galaxies out to a similar scale.
With an observed sample of 160 \oVI\ absorbers and 
over 50,000 galaxies at $z < 1$ and with \EAGLE, 
\citet{Finn2016} showed 
that \oVI\ has a very small velocity dispersion of 
$\lesssim100$\kms\ on $\sim$pMpc scale.
They also measured a lower correlation amplitude 
in the \oVI--galaxy cross-correlation
function than that from the galaxy autocorrelation function.
They concluded that this potentially implies that
\oVI\ and galaxies do not trace the same underlying 
matter distribution, 
and not all detected \oVI\ is close to the galaxies.
In fact, with the \FIREtwo\ cosmological simulations,
\citet{Hafen2020} showed that especially for low-mass galaxies,
most of the CGM at $z=2$ is accreted onto galaxies 
but ended up being ejected to the intergalactic medium by $z=0$.
This picture implies the transport of metals to 
the intergalactic scale at low-redshift. 
Altogether, these results 
indicate the possibility that a non-negligible 
amount of detected \oVI\ traces 
the warm-hot intergalactic medium (WHIM)
beyond the galaxy and group halo scales.  
The small velocity dispersion and LOS velocity also imply
a lack of substantial large-scale \oVI\ inflow 
and outflow at $\sim$pMpc scale,
which thereby suggest an early chemical enrichment history
of the WHIM
(e.g., \citealt{Wiersma2010}, and see
\citealt{Finn2016} for further discussion).

\subsection{
Caveat on identifying circumgalactic gas in observational analysis
and the \oVI\ comparison with simulations
}
\label{ssec:tosim}

Our results highlight the challenges
and limitations for observers to interpret
circumgalactic absorption measurements.
Figure~\ref{fig:o6_coldens_mhalo} clearly shows that
the gas within \rvir\ always produces a lower \oVI\ column
density compared to gas selected by the 
LOS velocity windows of $\pm300$\kms\ and $\pm500$\kms.
Even though the two velocity windows
produce indistinguishable \oVI\ column density measurements,
this does not imply the gas lies close to ($<$\rvir)
and/or is bound to the galaxies
(see also Figures~\ref{fig:o6_coldens} and 
\ref{fig:o6_coldens_4mass}).
In other words,
even if different \oVI\ velocity components 
within the \deltavlos\ window have comparable LOS velocities,
the detected gas possibly comes from distinct regions
and have different physical origins.
These velocity components may even be blended and 
indistinguishable depending on the LOS velocity separation 
and the resolution of the absorption spectum.
This potentially also explains why \oVI\ does not always
have matching velocity components with 
the centrally concentrated LIS gas \citep[e.g.,][]{Werk2016}.
Hence, the lack of knowledge 
of where the gas lies along LOS
makes it challenging for observers to interpret the \oVI\ measurements
and deduce the origin(s) of the detected \oVI\ gas.

Our analysis demonstrates a caveat for comparing 
\oVI\ measurements between observations and simulations.
With simulations, 
the column density is typically calculated by 
integrating the gas along a column with a specified path length
\citep[][etc.]{Hummels2013,Oppenheimer2016}.
In other words, the gas is selected based on its 3D position
relative to the galaxy.
This approach is different from observational analyses,
which identify the gas 
based on the LOS velocity regardless of the 3D location.

By selecting the gas around galaxies using the
LOS velocity windows,  
we still underpredict the \oVI\ column density 
compared to that observed
around $\sim$\lstar\ star-forming galaxies 
with $10^{14.5}$\percmsq\ at $b \lesssim$ 150 pkpc
\citep{Tumlinson2011}.  
However, the difference we find between velocity
and LOS distance selection partially accounts for 
the differences that have been reported between simulations
and observations and thereby 
reduces the discrepancies reported between the two.
Therefore, our analysis 
emphasizes the importance of recognizing 
how the use of different criteria 
for identifying the circumgalactic gas 
can lead to discrepancies in \oVI\ measurements 
between observations and cosmological simulations.
Zoom-in simulations may not also have 
a large enough volume 
to include all the gas within the velocity window
and may thereby underestimate the projection effects.

\vspace{2em}

We thank the referee for the detailed and thoughtful
comments that improved the manuscript.  We also
thank Hsiao-Wen Chen, Fakhri Zahedy,
and Tom Cooper for insightful discussions.
This work is partly funded by Vici grant 639.043.409 
from the Dutch Research Council (NWO)
and is partly supported in part
by the National Science Foundation under AST-1817125.
This work used the DiRAC@Durham facility 
managed by the Institute for Computational Cosmology 
on behalf of the STFC DiRAC HPC Facility
(www.dirac.ac.uk). 
The equipment was funded by BEIS capital funding 
via STFC capital grants ST/K00042X/1, ST/P002293/1,
ST/R002371/1 and ST/S002502/1, Durham University and 
STFC operations grant ST/R000832/1. 
DiRAC is part of the National e-Infrastructure.
%

\bibliography{master_eagle21_sim}

\end{document}

%% file: setdef.tex
%
%
\def\etal{{\rm et al.}}
\def\etali{{\it et al.\thinspace}}
\def\etns{{\rm et\thinspace al.}}   
\def\etaln{et al.\thinspace}

\def\EAGLE{\texttt{EAGLE}}
\def\AREPO{\texttt{AREPO}}
\def\ENZO{\texttt{ENZO}}
\def\GADGET{\texttt{GADGET-3}}
\def\OWLS{\texttt{OWLS}}
\def\FIRE{\texttt{FIRE}}
\def\FIREtwo{\texttt{FIRE-2}}
\def\illustris{\texttt{Illustris}}
\def\illustrisTNG{\texttt{IllustrisTNG}}

\def\mgIIdb{\ion{Mg}{2} $\lambda\lambda$2796, 2803}
\def\mgIIdbl{\ion{Mg}{2} $\lambda$2796}
\def\mgIIdbu{\ion{Mg}{2} $\lambda$2803}
\def\oI{[\ion{O}{1}] $\lambda$6300}
\def\oVIdb{\ion{O}{6} $\lambda\lambda$1032, 1038}
\def\sII{[\ion{S}{2}] $\lambda\lambda$6716, 6731}
\def\oIII{[\ion{O}{3}] $\lambda$5007}
\def\hI{\mbox {\ion{H}{1}}}
\def\mgII{\mbox {\ion{Mg}{2}}}
\def\nII{\mbox [{\ion{N}{2}}]}
\def\oVI{\mbox {\ion{O}{6}}}
\def\oVII{\mbox {\ion{O}{7}}}
\def\oVIII{\mbox {\ion{O}{8}}}
\def\cIV{\mbox {\ion{C}{4}}}
\def\halpha{$\mathrm{H}\alpha$}
\def\heII{\mbox {\ion{He}{2}}}

\def\mgplus{$\textrm{Mg}^{+}$}
\def\oVIion{O$^\mathrm{+5}$}

\def\kms{\mbox{km s$^{-1}$}}
\def\kmsMpc{\mbox{km s$^{-1}$ Mpc$^{-1}$}}
\def\kmstb{km s$^{-1}$}
\def\micron{\mbox{$\mu$m}}
\def\modotyr{\mbox {$\rm M_\odot$~yr$^{-1}$}}
\def\msununit{$\rm M_\odot$}
\def\percmsq{\mbox{cm$^{-2}$}}
\def\percmcube{\mbox{cm$^{-3}$}}

\def\rvir{$r_\mathrm{vir}$}
\def\mvir{$M_\mathrm{vir}$}
\def\mstar{$M_\star$}
\def\brvir{$b/r_\mathrm{vir}$}

\def\logmstar{$\log M_\star$}
\def\logmstarmsun{$\log (M_\star/\mathrm{M_\odot})$}
\def\logmvirmsun{$\log (M_\mathrm{vir}/\mathrm{M_\odot})$}

\def\NmgII{$N_\mathrm{MgII}$}

\def\deltavlos{$|\Delta v_\mathrm{LOS}|$}
\def\fmgiimis{$f_\mathrm{MgII,mis}$}
\def\fmgiimisb{$f_\mathrm{MgII,mis}(b)$}
\def\halovc{$v_\mathrm{c,halo}$}
\def\jstar{$\bm{j_\star}$}
\def\jstarscalar{$j_\star$}
\def\fcorotall{$F^\mathrm{corot+det}_\mathrm{all}$}
\def\fcorotdet{$f^\mathrm{corot+det}_\mathrm{det}$}
\def\logNmgii{log N(Mg\thinspace II)}
\def\logNovi{log N(O\thinspace VI)}

%
%
%
\def \dlow {\mbox{$400 {\rm ~l~mm}^{-1}$}}
\def \dhigh {\mbox{$600 {\rm ~l~mm}^{-1}$}}
\newcommand{\be}{\begin{equation}} \newcommand{\ba}{\begin{eqnarray}}
\newcommand{\ee}{\end{equation}} \newcommand{\ea}{\end{eqnarray}}
\def\-{{\em{---}}}
\def \mA {\mbox{${\rm m \AA} $} }
\def \rr {\mbox{${\rm RR}$} }
\def \rarb {\mbox{${\rm R_AR_B}$} }
\def \rara {\mbox{${\rm R_AR_A}$} }
\def \dd {\mbox{${\rm DD}$} }
\def \dada {\mbox{${\rm D_AD_A}$} }
\def \dadb {\mbox{${\rm D_AD_B}$} }
\def \dr {\mbox{${\rm DR}$} }
\def \darb {\mbox{${\rm D_AR_B}$} }
\def \dara {\mbox{${\rm D_AR_A}$} }
\def \dbra {\mbox{${\rm D_BR_A}$} }
\def \hMpc      {h^{-1}{\rm\ Mpc}}
\def \hkpc      {h^{-1}{\rm\ kpc}}
\def \h         {\hbox{$\, h$} }
\def \hinv      {\hbox{$\, h^{-1}$} }
\def \hinvseven    {\hbox{$\, h_{70}^{-1}$} }
\def\ewr{\mbox {EW$_r$}}
\def\ewo{\mbox {EW$_o$}}
\def\H7{\mbox {$h_{0.7}$}}
\def\naI{\mbox {\ion{Na}{1}}}
\def\mgI{\mbox {\ion{Mg}{1}}}
\def\feI{\mbox {\ion{Fe}{1}}}
\def\znII{\mbox {\ion{Zn}{2}}}
\def\crII{\mbox {\ion{Cr}{2}}}
\def\alI{\mbox {\sc Al~I~}}
\def\alII{\mbox {\sc Al~II~}}
\def\alIII{\mbox {\sc Al~III~}}
\def\mnII{\mbox {\ion{Mn}{2}}}
\def\niII{\mbox {\ion{Ni}{2}}}
\def\feII{\mbox {\ion{Fe}{2}}}
\def\feIII{\mbox {\ion{Fe}{3}}}
\def\sV{\mbox {\ion{S}{5}}}
\def\siIV{\mbox {\ion{Si}{4}}}
\def\siIII{\mbox {\ion{Si}{3}}}
\def\siII{\mbox {\ion{Si}{2}}}
\def\siI{\mbox {\ion{Si}{1}}}
\def\cII{\mbox {\ion{C}{2}}}
\def\cIII{\mbox {\ion{C}{3}}}
\def\llambda{\mbox {$\lambda$}}
\def\hlen{\mbox {$h_{0.7}^{-1}$}}
\def\lstarlya{\mbox {$L^*_{Ly\alpha}$}}
\def\IZw18{I~Zw~18}
\def\m82{M82}
\def\Ab{Abell~}
\def\gi{\mbox {\rm g-i}}
\def\ug{\mbox {\rm u-g}}
\def\br{\mbox {\rm b-r}}
\def\eqn{equation}
\def\vesc{\mbox {$v_{\rm esc}$}}
\def\heha{\mbox {He~I~$\lambda 5876$ / H$\alpha$}}
\def\xhe{\mbox {$\chi({\rm He}) / \chi({\rm H})$} }
\def\he{\mbox {\rm He}}
\def\hii{\mbox {${\rm H}^+$}}
\def\h{\mbox {\rm H}}
\def\mab{\mbox {$\rm m_{AB}$}}
\def\ssp{\baselineskip=13pt plus 1pt minus 1pt}
\def\tsp{\baselineskip=5pt plus 1pt minus 1pt}
%
%
\def\deg{\mbox {$^{\circ}$}}
\def\msun{\mbox {${\rm ~M_\odot}$}}
\def\zsun{\mbox {${\rm ~Z_{\odot}}$}}
\def\lsun{\mbox {${~\rm L_\odot}$}}
\def\msunyr{\mbox {$~{\rm M_\odot}$~yr$^{-1}$}}
\def\angs{\mbox {~\AA}}
\def\lya{\mbox {Ly$\alpha$}}
\def\lyb{\mbox {Ly$\beta$}}
\def\Ha{\mbox {H$\alpha$}}
\def\Hb{\mbox {H$\beta$}}
\def\Hg{\mbox {H$\gamma$}}
\def\tion{\mbox {$T_{\rm ion}$~}}
\def\ch{\mbox {$\bigtriangleup$}}
\def\grad{\mbox {$\bigtriangledown$}}
\def\lstar{\mbox {$L^*$}}
\def\line{\mbox {~$\lambda$}}
\def\lines{\mbox {~$\lambda\lambda$~}}
\def\h0{\mbox {~H$_0$}}
\def\q0{\mbox {~q$_0$}}
%
%
\def\auroral{[OIII]~$\lambda4363$~}
\def\auroral{[OIII]~$\lambda4363$~}
\def\ohsun{\mbox {(O/H)$_{\odot}$~}}

\def\O1ha{[OI]$\lambda6300$~/~H$\alpha$~}
\def\Ru{[OII]$\lambda\lambda3727$~/~[OIII]$\lambda5007$~}
\def\s2ha{[SII]$\lambda\lambda6717,31$~/~H$\alpha$~}
\def\2z2{HeII~$\lambda4686$~}
\def\z7{[NII]~$\lambda6583$ }
\def\N2{[NII]~$\lambda6583$~/~H$\alpha$~}
\def\16z2{[SII]~$\lambda\lambda6717, 6731$ }
\def\HgI{HgI~$\lambda4358$~}
\def\Sdensity{[SII]~$\lambda6717 / \lambda6731$}
\def\Temp{[OIII]~$\lambda\lambda4959 + 5007 ~{\rm to}~ \lambda4363$~}
%
%
\def\j{J}
\def\n{NGC~}
\def\asec{\ifmmode {'' }\else $''~$\fi}  
\def\amin{\ifmmode {' }\else $'~$\fi}    
\def\arcsper{\ifmmode \rlap.{'' }\else $\rlap{.}'' $\fi} 
\def\arcmper{\ifmmode \rlap.{' }\else $\rlap{.}' $\fi} 
\def\sles{\lesssim}
\def\sgreat{\gtrsim}
%
%
\def\gapp{\mbox {$_>\atop{^\sim}$}}  
\def\lapp{\mbox {$_<\atop{^\sim}$}}  
%
\def\kms{\mbox {~km~s$^{-1}$}}
\def\ergsec{~ergs~s$^{-1}$~}
\def\sb{~ergs~s$^{-1}$~cm$^{-2}$~arcsec$^{-2}$}
\def\flux{~ergs~s$^{-1}$~cm$^{-2}$}
\def\flam{~ergs~s$^{-1}$~cm$^{-2}$ \AA$^{-1}$}
\def\cm3{~cm$^{-3}$}
\def\col{\mbox {~cm$^{-2}$}}
\def\mpc3{~Mpc$^{3}$}
\def\mpc-3{~Mpc$^{-3}$}
\def\rate{~sec$~{-1}$}
\def\um{~${\mu}$m~}
\def\fig{{Figure}}
\def\figs{{Figures}}
\def\tbl{{Table}~}
\def\sec{{Sec.}~}
\def\x{{X-ray}~}
\def\xs{{X-rays}~}
\def\X{{X-Ray}~}

%
\def\et{{\rm et\thinspace al.}\ }   
\def\ets{{\rm et\thinspace al.'s}\ }   
\def\reff{\par\noindent\parskip=1pt\hangindent=3pc\hangafter=1}
%
%

%
\def\beginrefs{
         {\normalsize}
         {\noindent}
         \small
        \baselineskip=11pt
        \parindent=0pt
        \frenchspacing
        \parskip=1pt plus 1pt
        \everypar={\hangindent=0.42in}}

%% file: ms_x1.bbl
\begin{thebibliography}{}
\expandafter\ifx\csname natexlab\endcsname\relax\def\natexlab#1{#1}\fi

\bibitem[{{Appleby} {et~al.}(2021){Appleby}, {Dav{\'e}}, {Sorini},
  {Storey-Fisher}, \& {Smith}}]{Appleby2021}
{Appleby}, S., {Dav{\'e}}, R., {Sorini}, D., {Storey-Fisher}, K., \& {Smith},
  B. 2021, arXiv e-prints, arXiv:2102.10126

\bibitem[{{Bah{\'e}} {et~al.}(2016){Bah{\'e}}, {Crain}, {Kauffmann}, {Bower},
  {Schaye}, {Furlong}, {Lagos}, {Schaller}, {Trayford}, {Dalla Vecchia}, \&
  {Theuns}}]{Bahe2016}
{Bah{\'e}}, Y.~M., {Crain}, R.~A., {Kauffmann}, G., {et~al.} 2016, \mnras, 456,
  1115

\bibitem[{{Beckett} {et~al.}(2021){Beckett}, {Morris}, {Fumagalli}, {Bielby},
  {Tejos}, {Schaye}, {Jannuzi}, \& {Cantalupo}}]{Beckett2021}
{Beckett}, A., {Morris}, S., {Fumagalli}, M., {et~al.} 2021, arXiv e-prints,
  arXiv:2106.06416

\bibitem[{{Bertone} {et~al.}(2010{\natexlab{a}}){Bertone}, {Schaye}, {Booth},
  {Dalla Vecchia}, {Theuns}, \& {Wiersma}}]{Bertone2010a}
{Bertone}, S., {Schaye}, J., {Booth}, C.~M., {et~al.} 2010{\natexlab{a}},
  \mnras, 408, 1120

\bibitem[{{Bertone} {et~al.}(2010{\natexlab{b}}){Bertone}, {Schaye}, {Dalla
  Vecchia}, {Booth}, {Theuns}, \& {Wiersma}}]{Bertone2010b}
{Bertone}, S., {Schaye}, J., {Dalla Vecchia}, C., {et~al.} 2010{\natexlab{b}},
  \mnras, 407, 544

\bibitem[{{Bryan} \& {Norman}(1998)}]{BryanNorman1998}
{Bryan}, G.~L., \& {Norman}, M.~L. 1998, \apj, 495, 80

\bibitem[{{Chen} {et~al.}(2010){Chen}, {Helsby}, {Gauthier}, {Shectman},
  {Thompson}, \& {Tinker}}]{Chen2010}
{Chen}, H.-W., {Helsby}, J.~E., {Gauthier}, J.-R., {et~al.} 2010, \apj, 714,
  1521

\bibitem[{{Crain} {et~al.}(2015){Crain}, {Schaye}, {Bower}, {Furlong},
  {Schaller}, {Theuns}, {Dalla Vecchia}, {Frenk}, {McCarthy}, {Helly},
  {Jenkins}, {Rosas-Guevara}, {White}, \& {Trayford}}]{Crain2015}
{Crain}, R.~A., {Schaye}, J., {Bower}, R.~G., {et~al.} 2015, \mnras, 450, 1937

\bibitem[{{Crain} {et~al.}(2017){Crain}, {Bah{\'e}}, {Lagos}, {Rahmati},
  {Schaye}, {McCarthy}, {Marasco}, {Bower}, {Schaller}, {Theuns}, \& {van der
  Hulst}}]{Crain2017}
{Crain}, R.~A., {Bah{\'e}}, Y.~M., {Lagos}, C.~d.~P., {et~al.} 2017, \mnras,
  464, 4204

\bibitem[{{Davies} {et~al.}(2020){Davies}, {Crain}, {Oppenheimer}, \&
  {Schaye}}]{Davies2020}
{Davies}, J.~J., {Crain}, R.~A., {Oppenheimer}, B.~D., \& {Schaye}, J. 2020,
  \mnras, 491, 4462

\bibitem[{{Dolag} {et~al.}(2009){Dolag}, {Borgani}, {Murante}, \&
  {Springel}}]{Dolag2009}
{Dolag}, K., {Borgani}, S., {Murante}, G., \& {Springel}, V. 2009, \mnras, 399,
  497

\bibitem[{{Faucher-Gigu{\`e}re}(2020)}]{FaucherGiguere2020}
{Faucher-Gigu{\`e}re}, C.-A. 2020, \mnras, 493, 1614

\bibitem[{{Finn} {et~al.}(2016){Finn}, {Morris}, {Tejos}, {Crighton}, {Perry},
  {Fumagalli}, {Bielby}, {Theuns}, {Schaye}, {Shanks}, {Liske}, {Gunawardhana},
  \& {Bartle}}]{Finn2016}
{Finn}, C.~W., {Morris}, S.~L., {Tejos}, N., {et~al.} 2016, \mnras, 460, 590

\bibitem[{{Ford} {et~al.}(2016){Ford}, {Werk}, {Dav{\'e}}, {Tumlinson},
  {Bordoloi}, {Katz}, {Kollmeier}, {Oppenheimer}, {Peeples}, {Prochaska}, \&
  {Weinberg}}]{Ford2016}
{Ford}, A.~B., {Werk}, J.~K., {Dav{\'e}}, R., {et~al.} 2016, \mnras, 459, 1745

\bibitem[{{Furlong} {et~al.}(2015){Furlong}, {Bower}, {Theuns}, {Schaye},
  {Crain}, {Schaller}, {Dalla Vecchia}, {Frenk}, {McCarthy}, {Helly},
  {Jenkins}, \& {Rosas-Guevara}}]{Furlong2015}
{Furlong}, M., {Bower}, R.~G., {Theuns}, T., {et~al.} 2015, \mnras, 450, 4486

\bibitem[{{Furlong} {et~al.}(2017){Furlong}, {Bower}, {Crain}, {Schaye},
  {Theuns}, {Trayford}, {Qu}, {Schaller}, {Berthet}, \& {Helly}}]{Furlong2017}
{Furlong}, M., {Bower}, R.~G., {Crain}, R.~A., {et~al.} 2017, \mnras, 465, 722

\bibitem[{{Gutcke} {et~al.}(2017){Gutcke}, {Stinson}, {Macci{\`o}}, {Wang}, \&
  {Dutton}}]{Gutcke2017}
{Gutcke}, T.~A., {Stinson}, G.~S., {Macci{\`o}}, A.~V., {Wang}, L., \&
  {Dutton}, A.~A. 2017, \mnras, 464, 2796

\bibitem[{{Haardt} \& {Madau}(2001)}]{HaardtMadau2001}
{Haardt}, F., \& {Madau}, P. 2001, in Clusters of Galaxies and the High
  Redshift Universe Observed in X-rays, ed. D.~M. {Neumann} \& J.~T.~V. {Tran},
  64

\bibitem[{{Hafen} {et~al.}(2020){Hafen}, {Faucher-Gigu{\`e}re},
  {Angl{\'e}s-Alc{\'a}zar}, {Stern}, {Kere{\v{s}}}, {Esmerian}, {Wetzel},
  {El-Badry}, {Chan}, \& {Murray}}]{Hafen2020}
{Hafen}, Z., {Faucher-Gigu{\`e}re}, C.-A., {Angl{\'e}s-Alc{\'a}zar}, D.,
  {et~al.} 2020, \mnras, 494, 3581

\bibitem[{{Ho} \& {Martin}(2020)}]{HoMartin2020}
{Ho}, S.~H., \& {Martin}, C.~L. 2020, \apj, 888, 14

\bibitem[{{Ho} {et~al.}(2017){Ho}, {Martin}, {Kacprzak}, \&
  {Churchill}}]{Ho2017}
{Ho}, S.~H., {Martin}, C.~L., {Kacprzak}, G.~G., \& {Churchill}, C.~W. 2017,
  \apj, 835, 267

\bibitem[{{Ho} {et~al.}(2020){Ho}, {Martin}, \& {Schaye}}]{Ho2020}
{Ho}, S.~H., {Martin}, C.~L., \& {Schaye}, J. 2020, \apj, 904, 76

\bibitem[{{Huang} {et~al.}(2021){Huang}, {Chen}, {Shectman}, {Johnson},
  {Zahedy}, {Helsby}, {Gauthier}, \& {Thompson}}]{Huang2021}
{Huang}, Y.-H., {Chen}, H.-W., {Shectman}, S.~A., {et~al.} 2021, \mnras, 502,
  4743

\bibitem[{{Hummels} {et~al.}(2013){Hummels}, {Bryan}, {Smith}, \&
  {Turk}}]{Hummels2013}
{Hummels}, C.~B., {Bryan}, G.~L., {Smith}, B.~D., \& {Turk}, M.~J. 2013,
  \mnras, 430, 1548

\bibitem[{{Ji} {et~al.}(2020){Ji}, {Chan}, {Hummels}, {Hopkins}, {Stern},
  {Kere{\v{s}}}, {Quataert}, {Faucher-Gigu{\`e}re}, \& {Murray}}]{Ji2020}
{Ji}, S., {Chan}, T.~K., {Hummels}, C.~B., {et~al.} 2020, \mnras, 496, 4221

\bibitem[{{Johnson} {et~al.}(2015){Johnson}, {Chen}, \&
  {Mulchaey}}]{Johnson2015}
{Johnson}, S.~D., {Chen}, H.-W., \& {Mulchaey}, J.~S. 2015, \mnras, 449, 3263

\bibitem[{{Johnson} {et~al.}(2017){Johnson}, {Chen}, {Mulchaey}, {Schaye}, \&
  {Straka}}]{Johnson2017}
{Johnson}, S.~D., {Chen}, H.-W., {Mulchaey}, J.~S., {Schaye}, J., \& {Straka},
  L.~A. 2017, \apjl, 850, L10

\bibitem[{{Kacprzak} {et~al.}(2012){Kacprzak}, {Churchill}, \&
  {Nielsen}}]{Kacprzak2012}
{Kacprzak}, G.~G., {Churchill}, C.~W., \& {Nielsen}, N.~M. 2012, \apjl, 760, L7

\bibitem[{{Kacprzak} {et~al.}(2019){Kacprzak}, {Vander Vliet}, {Nielsen},
  {Muzahid}, {Pointon}, {Churchill}, {Ceverino}, {Arraki}, {Klypin},
  {Charlton}, \& {Lewis}}]{Kacprzak2019}
{Kacprzak}, G.~G., {Vander Vliet}, J.~R., {Nielsen}, N.~M., {et~al.} 2019,
  \apj, 870, 137

\bibitem[{{Lagos} {et~al.}(2015){Lagos}, {Crain}, {Schaye}, {Furlong}, {Frenk},
  {Bower}, {Schaller}, {Theuns}, {Trayford}, {Bah{\'e}}, \& {Dalla
  Vecchia}}]{Lagos2015}
{Lagos}, C.~d.~P., {Crain}, R.~A., {Schaye}, J., {et~al.} 2015, \mnras, 452,
  3815

\bibitem[{{Marra} {et~al.}(2021){Marra}, {Churchill}, {Kacprzak}, {Vander
  Vliet}, {Ceverino}, {Lewis}, {Nielsen}, {Muzahid}, \& {Charlton}}]{Marra2020}
{Marra}, R., {Churchill}, C.~W., {Kacprzak}, G.~G., {et~al.} 2021, \apj, 907, 8

\bibitem[{{Martin} {et~al.}(2019){Martin}, {Ho}, {Kacprzak}, \&
  {Churchill}}]{Martin2019}
{Martin}, C.~L., {Ho}, S.~H., {Kacprzak}, G.~G., \& {Churchill}, C.~W. 2019,
  \apj, 878, 84

\bibitem[{{McAlpine} {et~al.}(2016){McAlpine}, {Helly}, {Schaller}, {Trayford},
  {Qu}, {Furlong}, {Bower}, {Crain}, {Schaye}, {Theuns}, {Dalla Vecchia},
  {Frenk}, {McCarthy}, {Jenkins}, {Rosas-Guevara}, {White}, {Baes}, {Camps}, \&
  {Lemson}}]{McAlpine2016}
{McAlpine}, S., {Helly}, J.~C., {Schaller}, M., {et~al.} 2016, Astronomy and
  Computing, 15, 72

\bibitem[{{Moustakas} {et~al.}(2013){Moustakas}, {Coil}, {Aird}, {Blanton},
  {Cool}, {Eisenstein}, {Mendez}, {Wong}, {Zhu}, \& {Arnouts}}]{Moustakas2013}
{Moustakas}, J., {Coil}, A.~L., {Aird}, J., {et~al.} 2013, \apj, 767, 50

\bibitem[{{Nelson} {et~al.}(2018){Nelson}, {Kauffmann}, {Pillepich}, {Genel},
  {Springel}, {Pakmor}, {Hernquist}, {Weinberger}, {Torrey}, {Vogelsberger}, \&
  {Marinacci}}]{Nelson2018}
{Nelson}, D., {Kauffmann}, G., {Pillepich}, A., {et~al.} 2018, \mnras, 477, 450

\bibitem[{{Nielsen} {et~al.}(2015){Nielsen}, {Churchill}, {Kacprzak}, {Murphy},
  \& {Evans}}]{Nielsen2015}
{Nielsen}, N.~M., {Churchill}, C.~W., {Kacprzak}, G.~G., {Murphy}, M.~T., \&
  {Evans}, J.~L. 2015, \apj, 812, 83

\bibitem[{{Nielsen} {et~al.}(2017){Nielsen}, {Kacprzak}, {Muzahid},
  {Churchill}, {Murphy}, \& {Charlton}}]{Nielsen2017}
{Nielsen}, N.~M., {Kacprzak}, G.~G., {Muzahid}, S., {et~al.} 2017, \apj, 834,
  148

\bibitem[{{Oppenheimer} \& {Schaye}(2013)}]{OppenheimerSchaye2013}
{Oppenheimer}, B.~D., \& {Schaye}, J. 2013, \mnras, 434, 1063

\bibitem[{{Oppenheimer} {et~al.}(2018{\natexlab{a}}){Oppenheimer}, {Schaye},
  {Crain}, {Werk}, \& {Richings}}]{Oppenheimer2018lowion}
{Oppenheimer}, B.~D., {Schaye}, J., {Crain}, R.~A., {Werk}, J.~K., \&
  {Richings}, A.~J. 2018{\natexlab{a}}, \mnras, 481, 835

\bibitem[{{Oppenheimer} {et~al.}(2018{\natexlab{b}}){Oppenheimer}, {Segers},
  {Schaye}, {Richings}, \& {Crain}}]{Oppenheimer2018agn}
{Oppenheimer}, B.~D., {Segers}, M., {Schaye}, J., {Richings}, A.~J., \&
  {Crain}, R.~A. 2018{\natexlab{b}}, \mnras, 474, 4740

\bibitem[{{Oppenheimer} {et~al.}(2016){Oppenheimer}, {Crain}, {Schaye},
  {Rahmati}, {Richings}, {Trayford}, {Tumlinson}, {Bower}, {Schaller}, \&
  {Theuns}}]{Oppenheimer2016}
{Oppenheimer}, B.~D., {Crain}, R.~A., {Schaye}, J., {et~al.} 2016, \mnras, 460,
  2157

\bibitem[{{Planck Collaboration} {et~al.}(2014){Planck Collaboration}, {Ade},
  {Aghanim}, {Armitage-Caplan}, {Arnaud}, {Ashdown}, {Atrio-Barand ela},
  {Aumont}, {Baccigalupi}, {Banday}, {Barreiro}, {Bartlett}, {Battaner},
  {Benabed}, {Beno{\^\i}t}, {Benoit-L{\'e}vy}, {Bernard}, {Bersanelli},
  {Bielewicz}, {Bobin}, {Bock}, {Bonaldi}, {Bond}, {Borrill}, {Bouchet},
  {Bridges}, {Bucher}, {Burigana}, {Butler}, {Calabrese}, {Cappellini},
  {Cardoso}, {Catalano}, {Challinor}, {Chamballu}, {Chary}, {Chen}, {Chiang},
  {Chiang}, {Christensen}, {Church}, {Clements}, {Colombi}, {Colombo},
  {Couchot}, {Coulais}, {Crill}, {Curto}, {Cuttaia}, {Danese}, {Davies},
  {Davis}, {de Bernardis}, {de Rosa}, {de Zotti}, {Delabrouille}, {Delouis},
  {D{\'e}sert}, {Dickinson}, {Diego}, {Dolag}, {Dole}, {Donzelli}, {Dor{\'e}},
  {Douspis}, {Dunkley}, {Dupac}, {Efstathiou}, {Elsner}, {En{\ss}lin},
  {Eriksen}, {Finelli}, {Forni}, {Frailis}, {Fraisse}, {Franceschi}, {Gaier},
  {Galeotta}, {Galli}, {Ganga}, {Giard}, {Giardino}, {Giraud-H{\'e}raud},
  {Gjerl{\o}w}, {Gonz{\'a}lez-Nuevo}, {G{\'o}rski}, {Gratton}, {Gregorio},
  {Gruppuso}, {Gudmundsson}, {Haissinski}, {Hamann}, {Hansen}, {Hanson},
  {Harrison}, {Henrot-Versill{\'e}}, {Hern{\'a}ndez-Monteagudo}, {Herranz},
  {Hildebrand t}, {Hivon}, {Hobson}, {Holmes}, {Hornstrup}, {Hou}, {Hovest},
  {Huffenberger}, {Jaffe}, {Jaffe}, {Jewell}, {Jones}, {Juvela},
  {Keih{\"a}nen}, {Keskitalo}, {Kisner}, {Kneissl}, {Knoche}, {Knox}, {Kunz},
  {Kurki-Suonio}, {Lagache}, {L{\"a}hteenm{\"a}ki}, {Lamarre}, {Lasenby},
  {Lattanzi}, {Laureijs}, {Lawrence}, {Leach}, {Leahy}, {Leonardi},
  {Le{\'o}n-Tavares}, {Lesgourgues}, {Lewis}, {Liguori}, {Lilje},
  {Linden-V{\o}rnle}, {L{\'o}pez-Caniego}, {Lubin}, {Mac{\'\i}as-P{\'e}rez},
  {Maffei}, {Maino}, {Mand olesi}, {Maris}, {Marshall}, {Martin},
  {Mart{\'\i}nez-Gonz{\'a}lez}, {Masi}, {Massardi}, {Matarrese}, {Matthai},
  {Mazzotta}, {Meinhold}, {Melchiorri}, {Melin}, {Mendes}, {Menegoni},
  {Mennella}, {Migliaccio}, {Millea}, {Mitra}, {Miville-Desch{\^e}nes},
  {Moneti}, {Montier}, {Morgante}, {Mortlock}, {Moss}, {Munshi}, {Murphy},
  {Naselsky}, {Nati}, {Natoli}, {Netterfield}, {N{\o}rgaard-Nielsen},
  {Noviello}, {Novikov}, {Novikov}, {O'Dwyer}, {Osborne}, {Oxborrow}, {Paci},
  {Pagano}, {Pajot}, {Paladini}, {Paoletti}, {Partridge}, {Pasian},
  {Patanchon}, {Pearson}, {Pearson}, {Peiris}, {Perdereau}, {Perotto},
  {Perrotta}, {Pettorino}, {Piacentini}, {Piat}, {Pierpaoli}, {Pietrobon},
  {Plaszczynski}, {Platania}, {Pointecouteau}, {Polenta}, {Ponthieu}, {Popa},
  {Poutanen}, {Pratt}, {Pr{\'e}zeau}, {Prunet}, {Puget}, {Rachen}, {Reach},
  {Rebolo}, {Reinecke}, {Remazeilles}, {Renault}, {Ricciardi}, {Riller},
  {Ristorcelli}, {Rocha}, {Rosset}, {Roudier}, {Rowan-Robinson},
  {Rubi{\~n}o-Mart{\'\i}n}, {Rusholme}, {Sandri}, {Santos}, {Savelainen},
  {Savini}, {Scott}, {Seiffert}, {Shellard}, {Spencer}, {Starck}, {Stolyarov},
  {Stompor}, {Sudiwala}, {Sunyaev}, {Sureau}, {Sutton}, {Suur-Uski}, {Sygnet},
  {Tauber}, {Tavagnacco}, {Terenzi}, {Toffolatti}, {Tomasi}, {Tristram},
  {Tucci}, {Tuovinen}, {T{\"u}rler}, {Umana}, {Valenziano}, {Valiviita}, {Van
  Tent}, {Vielva}, {Villa}, {Vittorio}, {Wade}, {Wandelt}, {Wehus}, {White},
  {White}, {Wilkinson}, {Yvon}, {Zacchei}, \& {Zonca}}]{Planck2014}
{Planck Collaboration}, {Ade}, P.~A.~R., {Aghanim}, N., {et~al.} 2014, \aap,
  571, A16

\bibitem[{{Ploeckinger} \& {Schaye}(2020)}]{PloeckingerSchaye2020}
{Ploeckinger}, S., \& {Schaye}, J. 2020, \mnras, 497, 4857

\bibitem[{{Prochaska} {et~al.}(2011){Prochaska}, {Weiner}, {Chen}, {Mulchaey},
  \& {Cooksey}}]{Prochaska2011}
{Prochaska}, J.~X., {Weiner}, B., {Chen}, H.~W., {Mulchaey}, J., \& {Cooksey},
  K. 2011, \apj, 740, 91

\bibitem[{{Prochaska} {et~al.}(2019){Prochaska}, {Burchett}, {Tripp}, {Werk},
  {Willmer}, {Howk}, {Lange}, {Tejos}, {Meiring}, {Tumlinson}, {Lehner},
  {Ford}, \& {Dav{\'e}}}]{Prochaska2019}
{Prochaska}, J.~X., {Burchett}, J.~N., {Tripp}, T.~M., {et~al.} 2019, \apjs,
  243, 24

\bibitem[{{Rahmati} {et~al.}(2015){Rahmati}, {Schaye}, {Bower}, {Crain},
  {Furlong}, {Schaller}, \& {Theuns}}]{Rahmati2015}
{Rahmati}, A., {Schaye}, J., {Bower}, R.~G., {et~al.} 2015, \mnras, 452, 2034

\bibitem[{{Rahmati} {et~al.}(2016){Rahmati}, {Schaye}, {Crain}, {Oppenheimer},
  {Schaller}, \& {Theuns}}]{Rahmati2016}
{Rahmati}, A., {Schaye}, J., {Crain}, R.~A., {et~al.} 2016, \mnras, 459, 310

\bibitem[{{Roca-F{\`a}brega} {et~al.}(2019){Roca-F{\`a}brega}, {Dekel},
  {Faerman}, {Gnat}, {Strawn}, {Ceverino}, {Primack}, {Macci{\`o}}, {Dutton},
  {Prochaska}, \& {Stern}}]{RocaFabrega2019}
{Roca-F{\`a}brega}, S., {Dekel}, A., {Faerman}, Y., {et~al.} 2019, \mnras, 484,
  3625

\bibitem[{{Schaller} {et~al.}(2015){Schaller}, {Dalla Vecchia}, {Schaye},
  {Bower}, {Theuns}, {Crain}, {Furlong}, \& {McCarthy}}]{Schaller2015}
{Schaller}, M., {Dalla Vecchia}, C., {Schaye}, J., {et~al.} 2015, \mnras, 454,
  2277

\bibitem[{{Schaye} {et~al.}(2015){Schaye}, {Crain}, {Bower}, {Furlong},
  {Schaller}, {Theuns}, {Dalla Vecchia}, {Frenk}, {McCarthy}, {Helly},
  {Jenkins}, {Rosas-Guevara}, {White}, {Baes}, {Booth}, {Camps}, {Navarro},
  {Qu}, {Rahmati}, {Sawala}, {Thomas}, \& {Trayford}}]{Schaye2015}
{Schaye}, J., {Crain}, R.~A., {Bower}, R.~G., {et~al.} 2015, \mnras, 446, 521

\bibitem[{{Schroetter} {et~al.}(2019){Schroetter}, {Bouch{\'e}}, {Zabl},
  {Contini}, {Wendt}, {Schaye}, {Mitchell}, {Muzahid}, {Marino}, {Bacon},
  {Lilly}, {Richard}, \& {Wisotzki}}]{Schroetter2019}
{Schroetter}, I., {Bouch{\'e}}, N.~F., {Zabl}, J., {et~al.} 2019, \mnras, 490,
  4368

\bibitem[{{Springel}(2005)}]{Springel2005}
{Springel}, V. 2005, \mnras, 364, 1105

\bibitem[{{Springel} {et~al.}(2001){Springel}, {White}, {Tormen}, \&
  {Kauffmann}}]{Springel2001}
{Springel}, V., {White}, S.~D.~M., {Tormen}, G., \& {Kauffmann}, G. 2001,
  \mnras, 328, 726

\bibitem[{{Stern} {et~al.}(2018){Stern}, {Faucher-Gigu{\`e}re}, {Hennawi},
  {Hafen}, {Johnson}, \& {Fielding}}]{Stern2018}
{Stern}, J., {Faucher-Gigu{\`e}re}, C.-A., {Hennawi}, J.~F., {et~al.} 2018,
  \apj, 865, 91

\bibitem[{{Strawn} {et~al.}(2021){Strawn}, {Roca-F{\`a}brega}, {Mandelker},
  {Primack}, {Stern}, {Ceverino}, {Dekel}, {Wang}, \& {Dange}}]{Strawn2021}
{Strawn}, C., {Roca-F{\`a}brega}, S., {Mandelker}, N., {et~al.} 2021, \mnras,
  501, 4948

\bibitem[{{Suresh} {et~al.}(2017){Suresh}, {Rubin}, {Kannan}, {Werk},
  {Hernquist}, \& {Vogelsberger}}]{Suresh2017}
{Suresh}, J., {Rubin}, K. H.~R., {Kannan}, R., {et~al.} 2017, \mnras, 465, 2966

\bibitem[{{Trayford} {et~al.}(2015){Trayford}, {Theuns}, {Bower}, {Schaye},
  {Furlong}, {Schaller}, {Frenk}, {Crain}, {Dalla Vecchia}, \&
  {McCarthy}}]{Trayford2015}
{Trayford}, J.~W., {Theuns}, T., {Bower}, R.~G., {et~al.} 2015, \mnras, 452,
  2879

\bibitem[{{Trayford} {et~al.}(2017){Trayford}, {Camps}, {Theuns}, {Baes},
  {Bower}, {Crain}, {Gunawardhana}, {Schaller}, {Schaye}, \&
  {Frenk}}]{Trayford2017}
{Trayford}, J.~W., {Camps}, P., {Theuns}, T., {et~al.} 2017, \mnras, 470, 771

\bibitem[{{Tumlinson} {et~al.}(2017){Tumlinson}, {Peeples}, \&
  {Werk}}]{Tumlinson2017}
{Tumlinson}, J., {Peeples}, M.~S., \& {Werk}, J.~K. 2017, \araa, 55, 389

\bibitem[{{Tumlinson} {et~al.}(2011){Tumlinson}, {Thom}, {Werk}, {Prochaska},
  {Tripp}, {Weinberg}, {Peeples}, {O'Meara}, {Oppenheimer}, {Meiring}, {Katz},
  {Dav{\'e}}, {Ford}, \& {Sembach}}]{Tumlinson2011}
{Tumlinson}, J., {Thom}, C., {Werk}, J.~K., {et~al.} 2011, Science, 334, 948

\bibitem[{{Turner} {et~al.}(2017){Turner}, {Schaye}, {Crain}, {Rudie},
  {Steidel}, {Strom}, \& {Theuns}}]{Turner2017}
{Turner}, M.~L., {Schaye}, J., {Crain}, R.~A., {et~al.} 2017, \mnras, 471, 690

\bibitem[{{Turner} {et~al.}(2016){Turner}, {Schaye}, {Crain}, {Theuns}, \&
  {Wendt}}]{Turner2016}
{Turner}, M.~L., {Schaye}, J., {Crain}, R.~A., {Theuns}, T., \& {Wendt}, M.
  2016, \mnras, 462, 2440

\bibitem[{Wendland(1995)}]{Wendland1995}
Wendland, H. 1995, Advances in Computational Mathematics, 4, 389

\bibitem[{{Werk} {et~al.}(2016){Werk}, {Prochaska}, {Cantalupo}, {Fox},
  {Oppenheimer}, {Tumlinson}, {Tripp}, {Lehner}, \& {McQuinn}}]{Werk2016}
{Werk}, J.~K., {Prochaska}, J.~X., {Cantalupo}, S., {et~al.} 2016, \apj, 833,
  54

\bibitem[{{Wiersma} {et~al.}(2010){Wiersma}, {Schaye}, {Dalla Vecchia},
  {Booth}, {Theuns}, \& {Aguirre}}]{Wiersma2010}
{Wiersma}, R. P.~C., {Schaye}, J., {Dalla Vecchia}, C., {et~al.} 2010, \mnras,
  409, 132

\bibitem[{{Wiersma} {et~al.}(2009){Wiersma}, {Schaye}, \&
  {Smith}}]{Wiersma2009}
{Wiersma}, R. P.~C., {Schaye}, J., \& {Smith}, B.~D. 2009, \mnras, 393, 99

\bibitem[{{Wijers} {et~al.}(2020){Wijers}, {Schaye}, \&
  {Oppenheimer}}]{Wijers2020}
{Wijers}, N.~A., {Schaye}, J., \& {Oppenheimer}, B.~D. 2020, \mnras, 498, 574

\bibitem[{{Wijers} {et~al.}(2019){Wijers}, {Schaye}, {Oppenheimer}, {Crain}, \&
  {Nicastro}}]{Wijers2019}
{Wijers}, N.~A., {Schaye}, J., {Oppenheimer}, B.~D., {Crain}, R.~A., \&
  {Nicastro}, F. 2019, \mnras, 488, 2947

\bibitem[{{Zabl} {et~al.}(2019){Zabl}, {Bouch{\'e}}, {Schroetter}, {Wendt},
  {Finley}, {Schaye}, {Conseil}, {Contini}, {Marino}, {Mitchell}, {Muzahid},
  {Pezzulli}, \& {Wisotzki}}]{Zabl2019}
{Zabl}, J., {Bouch{\'e}}, N.~F., {Schroetter}, I., {et~al.} 2019, \mnras, 485,
  1961

\bibitem[{{Zahedy} {et~al.}(2019){Zahedy}, {Chen}, {Johnson}, {Pierce},
  {Rauch}, {Huang}, {Weiner}, \& {Gauthier}}]{Zahedy2019}
{Zahedy}, F.~S., {Chen}, H.-W., {Johnson}, S.~D., {et~al.} 2019, \mnras, 484,
  2257

\end{thebibliography}
